\documentclass{article}
\usepackage{graphicx,mathtools,natbib,amssymb,lineno,setspace}
\usepackage[margin=2.5cm]{geometry}
\usepackage[utf8]{inputenc}

\newcommand{\e}{\text{e}}
\newcommand{\xhat}{\widehat{\boldsymbol{x}}}
\newcommand{\zhat}{\widehat{\boldsymbol{z}}}
\newcommand{\pd}[2]{\frac{\partial{#1}}{\partial{#2}}}

\newcommand{\ld}[2]{\frac{\text{D}{#1}}{\text{D}{#2}}}
\newcommand{\pert}[1]{\Tilde{#1}}
\newcommand{\curvature}{\chi}
\newcommand{\deb}{\text{De}}
\newcommand{\yield}{M_\mathcal{Y}}
\newcommand{\yieldref}{\mathcal{M}} 
\newcommand{\ystress}{\sigma_\mathcal{Y}}
\newcommand{\weakcurvature}{\curvature_W}
\newcommand{\curvatureref}{\mathcal{X}_W}

\title{How sea level paces faulting\\ at fast-spreading mid-ocean ridges}
\author{Richard F.~Katz$^{1*}$, Peter Huybers$^2$ \\[2mm] 
    \small Affiliations: $^1$ Department of Earth Sciences, University of Oxford, Oxford UK, OX2 6NA;\\ \small $^2$ Department of Earth and Planetary Sciences, Cambridge, MA 02138; \\ \small $^*$ Corresponding author, richard.katz@earth.ox.ac.uk.}

\begin{document}

\maketitle


\begin{abstract}
    Abyssal hills---arguably the most extensive coherent pattern in Earth’s surface topography---record the spacing of normal faults formed at mid-ocean ridges. At fast-spreading ridges, spectral analysis of high-resolution bathymetry reveals a pronounced peak near 41~ky, coincident with obliquity-paced sea-level variability during the Pleistocene. The origin of this apparent imprint of orbital-cycle forcing on seafloor structure is unresolved. We propose that glacial--interglacial sea-level variability influences fault spacing through its effect on plate thickness and the resulting modulation of flexural stresses during plate unbending.

    Sea-level change modulates mantle melting rates and magma supply at ridge axes, generating variations in the properties of the accreting plate. As the plate moves off axis, it unbends from its ingrown curvature, producing tensile fibre stresses that drive normal faulting. We hypothesise that small perturbations in elastic plate thickness modulate these stresses and thereby influence the spacing of faults. To test this, we extend the elastic unbending theory of \cite{Buck2001} to incorporate (\textit{i}) spatially variable plate thickness and (\textit{ii}) yield-weakening viscoplastic flexure that allows deformation to localise into discrete kinks. We interpret these kinks as faults. 
    
    Linearised analysis shows that plate-thickness perturbations generate proportionate fibre-stress variations. Numerical solutions demonstrate that perturbations as small as $\sim$0.1\% can phase-lock faulting to the imposed forcing. When driven by plate-thickness perturbations derived from the Pleistocene $\delta^{18}$O record, the model predicts fault spacings concentrated near 41~ky in the early Pleistocene and near 100~ky in the late Pleistocene, consistent with observed abyssal-hill spectra. These results represent a quantitative mechanism by which small variations in plate thickness can transmit sea-level variability of glacial--interglacial cycles into tectonic structure, and provide a framework for fault-spacing selection in flexing plates.
\end{abstract}

\section{Introduction}

On the flanks of the global mid-ocean ridge (MOR) system, oceanic plates are corrugated by a progression of abyssal hills. These topographic features have a characteristic height of order 100~m and extend for hundreds of kilometres roughly parallel to the ridge \citep{menard1964marine}.  The spacing of hills varies, but is typically a few kilometres. Considering the rate of plate spreading, this interval also represents a spacing in time.  Using a global compilation of high-resolution data, \cite{Huybers2022} performed spectral analysis on bathymetric transects perpendicular to the ridge axis. For fast-spreading MORs, they found a significant concentration of spectral energy at periods of about 41~ky.  A similar concentration is found in the spectrum of eustatic sea level over the Pleistocene \citep{hays76} that is related to changes in Earth's obliquity \citep{imbrie1982}. It has been noted that sea-level variation on glacial--interglacial timescales drives fluctuations in MOR magma supply \citep{Huybers2009, Lund2011, Crowley2015}, but the relationship of abyssal hills to magma supply remains unclear. Abyssal hills are fault-bounded features with an origin that is dominantly tectonic \citep{Olive2015}, raising the question:  what mechanism would relate sea-level variations to the spacing of sea-floor faults?

Here we hypothesise that magma-supply variations driven by sea level cause a variation in the elastic thickness of the oceanic plate.  As the plate moves away from the topographic high of the ridge axis it unbends, flattening into the abyssal plains of the ocean basins \citep{Buck2001}. In this unbending process, plate thickness perturbations give rise to stress perturbations and these, in turn, pace faulting.  Faults---and their associated abyssal hills---therefore appear with a spacing corresponding to the 41~ky Milankovitch periodicity of sea-level variations.  In what follows, we quantify this hypothesis and demonstrate its physical plausibility and consistency with observations.

It is foundational to this hypothesis that sea-level variation can alter the rate of magma supply at mid-ocean ridges. This supply arises by near-isentropic decompression of steadily upwelling asthenosphere, which can be appreciably perturbed by variations in sea level.  For example, sea level rose at  $\sim$1~cm/a over $\sim$10~ka during the last deglaciation, giving a mantle pressurisation rate of $\sim$10~Pa/a.  This can be compared with a mantle upwelling of 3.5~cm/a that (accounting for mantle density) leads to a decompression rate of $\sim$100~Pa/a. The comparison implies that sea-level rise causes $\sim$10\% reductions in melt-production rate \citep{Huybers2009, Lund2011}.  This perturbation is admitted to the rate of ridge-axis magma supply only if it persists over a period that is comparable to the time-scale for melt extraction to the ridge axis \citep{Crowley2015, Cerpa2019}.  At fast-spreading ridges, the characteristic melt-extraction time roughly aligns with the 41~ka period of Pleistocene obliquity variations \citep{Katz2022}. This raises the questions of how such melt-supply variations might be expressed on the sea floor, and whether they can be observed.

Over the last decade, a number of studies have asked whether sea-floor topography of abyssal hills records the history of magma-supply variations with distance from the ridge axis, and hence time in the past.  Spectral peaks at periods that are prominent in Pleistocene variation in sea level were observed for select transects across the Atlantic--Antarctic Discord \citep{Crowley2015}, the East Pacific Rise \citep{Tolstoy2015kq}, and the Chile Rise \citep{Huybers2016}. \cite{Boulahanis2020gm} used seismic reflections to image variations in crustal thickness in the East Pacific Rise that are consistent with the 100~ky glacial--interglacial timescale. However, other ridge data fail to show these spectral peaks \citep{Goff2018}, raising questions regarding the significance and consistency of any mechanistic relationship between sea level and abyssal-hill spacing. To address this, \cite{Huybers2022} used high-resolution swath-bathymetry data from 17 ridge segments worldwide.  For each dataset, spectra were stacked over many transects, allowing confidence intervals to be assigned to spectral peaks. A robust inference is the presence of a $\sim$41~ka spectral peak at fast-spreading ridges (those with half-spreading rates of $U\ge3.8$~cm/a).  If this spectral result is taken as a fingerprint, what is the mechanism by which it was imprinted?

Mechanisms have been proposed and debated. \cite{Crowley2015} argued for a magmatic origin of abyssal-hill topography associated with isostatic balance at the ridge axis. \cite{Olive2015} disagreed, and instead argued for a tectonic origin. They highlighted observational evidence of faulting, and showed that bending stiffness inhibits isostatic compensation at the relevant wavelengths. Using a well-established numerical model in a limited parameter range, \cite{Olive2015} found that simulated fault spacing is unaffected by 100\% variations in magma supply.  However, using the same model with parameters outside this range, \cite{Huybers2022} found that the faults can be paced by such magma-supply variations. Nonetheless, the debate remains unresolved because of two major limitations in the application of this particular numerical model to magma-supply variations at fast spreading ridges. The first is that its magma-supply rate is treated as an on-off switch rather than being modulated by, for example, 10\% fluctuations.  The second issue is that the morphological regime of the simulated ridge axis is appropriate for slow-to-intermediate spreading rates.  In this regime, a magma-starved MOR is subjected to stretching by far-field tensile stress, and develops an axial valley. Extension is accommodated by normal faults that form where the elastic plate is youngest and thinnest---at the ridge axis---and dip inwards \citep{Buck1998hw}. This behaviour stands in contrast to fast-spreading ridges with robust magma supply \citep{Buck2005eo}.

At fast-spreading mid-ocean ridges, plate divergence is accommodated by the injection of abundant dikes at the ridge axis; only small-offset faults form there \citep{carbotte2006, escartin2007}. Larger normal-fault offsets accrue at distances of $\sim$30~km from the ridge axis, with dips that are toward and away from the axis in approximately equal number \citep{crowder2000, escartin2007}. \cite{Buck2001} and \cite{Shah2003} explained these observations and several others by a model, illustrated in Figure~\ref{fig:schematic}, where the tensile stresses for normal faulting arise by unbending of a neutrally curved plate. The ridge axis is an isostatic high due to the buoyancy of magma beneath; this buoyancy also drives dike emplacement. The plate's elastic thickness is established by vigorous hydrothermal circulation that cools the dikes. But the plate is not born flat---instead it inherits the curvature needed to connect the axial high to the distal plane.  This ingrown curvature does not generate flexural stress, it is elastically neutral. As the plate moves away from the axis and subsides into the isostatic balance of the abyssal plains, its curvature is eliminated. \cite{Buck2001} shows how this unbending of the ingrown, neutral curvature creates tensile stress in the upper half of the plate. These stresses arise at the same distance from the axis where normal-fault offsets are observed to increase.

Flexural stresses are commonly invoked to explain observed patterns of tectonic faulting and seismicity. At ultraslow-spreading ridges, for example, flexure along the detachment fault is associated with distinct patches of normal and reverse faulting \citep{parnell-turner2017, jackson2023}. At slow-spreading ridges, where an axial valley gives the plate a negative ingrown curvature, isostatic unbending leads to reverse slip on pre-existing normal faults that decreases their throw \citep[a.k.a.~``unfaulting,''][]{olive2024}. Flexural stress also drives faulting of oceanic plates as they bend into subduction zones \citep{masson1991fault}, which is the subject of various geodynamic models \citep{Turcotte2002, faccenda2009, zhou2015}.

Analytical models of plate flexure are based on classical Euler--Bernoulli beam theory \citep{Mansfield1989, Turcotte2002}; numerical models rely on this theory for their interpretation. The Euler-Bernoulli bending stiffness has a cubic dependence on plate thickness. Despite this nonlinear sensitivity, and although Earth is heterogeneous at any relevant scale, the implications of variation in plate thickness have seldom (if ever) been considered.  And yet if broad plate flexure is the cause of localised faulting, then even small plate-thickness variations could modulate the stresses that drive faulting, and hence control the pattern of faulting itself. The flexural analysis that we develop here supports this idea.

Our approach is to extend the model proposed by \cite{Buck2001} in two key ways. First, we incorporate faulting as an auto-localising viscoplastic process.  In our theory, flexure is elastic for bending moments $M$ that are smaller, in absolute value, than a critical yield moment $\yield$. Moments that exceed $\yield$ drive viscous flexure---an irreversible process that occurs rapidly if the viscosity is small.  Crucially, by  requiring $\yield$ to decrease with increasing viscous flexure, we represent the inherent self-weakening and localising tendency of fractures. The second extension of \cite{Buck2001}, illustrated in Figure~\ref{fig:schematic}, is to consider a non-uniform plate thickness $h$, and to associate the thickness perturbations with sea-level-driven variation in melt supply.  

Using numerical solutions and linearised analysis, we show that variations in plate thickness as small as $\sim$0.1\% are adequate to pace faulting. Models forced by reconstructed Pleistocene sea level (via its assumed effect on $h$) develop a pattern of faulting that reflects the spectrum of melt-supply variations, even in the presence of substantial additive noise. We interpret these results as establishing that plate-thickness variability provides a physically plausible pathway by which sea-level forcing can influence fault spacing, but not that it uniquely explains the observed spectral signal.

In section~\ref{sec:theory} we develop the rheological model (\S\ref{sec:theory-rheology}) and the force balance (\S\ref{sec:theory-forces}), and give estimates for dimensional and dimensionless parameters (\S\ref{sec:theory-parameters}). We then present results in section~\ref{sec:results}. These begin with a demonstration of how the theory models faults as viscoplastic kinks in a plate of uniform thickness (\S\ref{sec:periodic-kinks}). We then consider linearised analysis (\S\ref{sec:results-linearised}) and numerical solution (\S\ref{sec:results-monochrome-pacing}) for a plate with monochromatic thickness perturbations, demonstrating how this paces kink--faults.  Section~\ref{sec:results-sea-level} applies the theory to cases where plate-thickness perturbations are forced by Pleistocene sea level.  A discussion of the results, their interpretation, and the limitations of our approach follows in section~\ref{sec:discussion}. A brief, concluding summary is given in \S\ref{sec:conclusion}.

\section{The theory}
\label{sec:theory}

\begin{figure}
    \centering
    \includegraphics[width=0.8\linewidth]{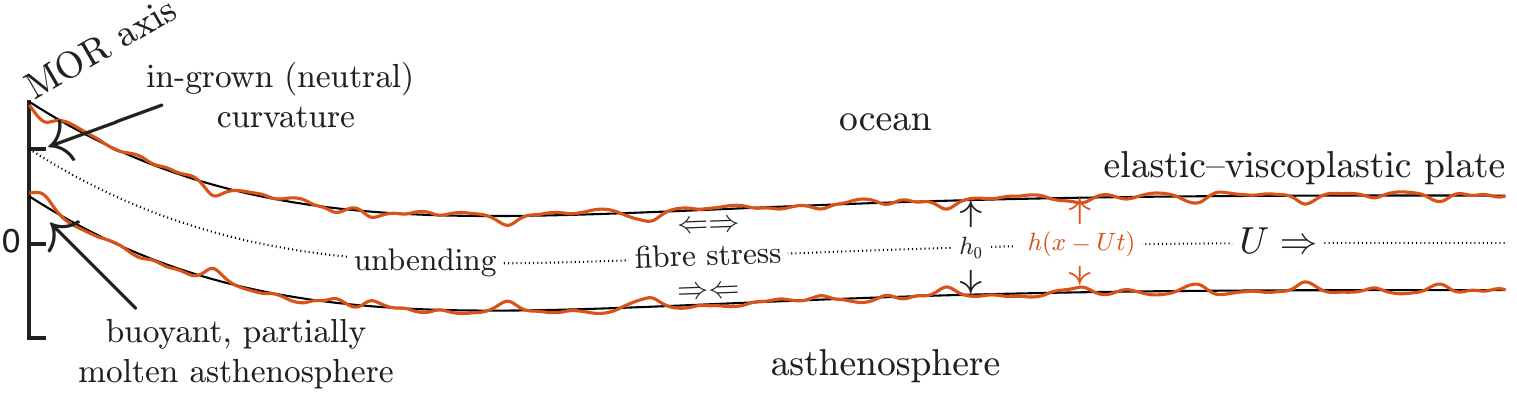}
    \caption{Schematic diagram (not to scale) of the elastic--viscoplastic plate. The $z$ direction is upward, opposite gravity. The $x$ direction is to the right, with the MOR axis at $x=0$. The plate is continuously created at the ridge axis and moves to the right at speed $U$. With distance from the ridge axis, the plate tends toward isostatic balance. This unbends the plate's ingrown curvature and creates fibre stresses that are relieved by normal faulting. The total thickness of the plate $h$ is perturbed away from the uniform thickness $h_0$ that was considered by \cite{Buck2001}. }
    \label{fig:schematic}
\end{figure}

We consider a semi-infinite plate of density $\rho$, with its boundary at $x=0$ representing the mid-ocean ridge. Upward displacement of the plate is measured in the $z$ direction and gravity acts downward, $\boldsymbol{g} = -g\zhat$.  The plate sits above a mantle of equal density and beneath an ocean of density $\rho_o$. Both the mantle and the ocean are treated as quasi-static.  A schematic diagram of the plate is shown in Figure~\ref{fig:schematic}. 

Distinct from \cite{Buck2001}, we assume that the plate thickness $h$ is not uniform, but rather has variations that are created by time-dependence of the melt supply at the ridge axis. The thickness is separated into its mean and variations as
\begin{equation}
    \label{eq:plate-thickness-decomp}
    h(x,t) = h_0 + \pert{h}(x,t),
\end{equation}
where $\Tilde{}$ represents a field with zero mean.

We assume cylindrical bending of the plate such that there is no variation of any quantity in the $y$ direction. Bending is described by a one-dimensional field of vertical displacements $z=w(x,t)$ of the plate's centre-line. We assume that this vertical displacement and its gradient $\partial_x w$ are sufficiently small that our horizontal coordinate $x$ is an adequate approximation for lateral variations of the plate.  With this approximation, plate motion has a velocity $U\xhat$, where the speed $U$ is uniform and constant. Furthermore, the curvature of the plate $\curvature(x,t)$ is the second derivative of vertical displacement,
\begin{equation}
    \label{eq:curvature-definition}
    \curvature(x,t) \equiv \partial_{xx}w.
\end{equation}
For concision, we express partial derivatives with respect to $x$ with a prime ($'$) and those with respect to time with a dot ($\dot{\,}$). In this notation, $\curvature \equiv w''.$

Our strategy in developing the model is to relate the bending moment $M(x,t)$ to the local curvature $\curvature(x,t)$, and hence to the vertical displacement $w(x,t)$.  This relationship must capture the effect of imposed variations in the thickness of the plate $\pert{h}(x,t)$. Plate-thickness variations are frozen into the plate; they are transported at velocity $U\xhat$ and are hence of the form $\pert{h}(x-Ut)$, but do not otherwise evolve.

\subsection{Elastic--viscoplastic flexure with yield weakening}
\label{sec:theory-rheology}

We formulate the rheological hypothesis directly in terms of flexure, relating the bending moment to the plate curvature. We assume a Maxwell-type combination of elastic and plastic deformation, such that the total curvature satisfies
\begin{equation}
    \label{eq:maxwell-total-curvature}
    \curvature = \curvature_e + \curvature_p,
\end{equation}
where $_e$ and $_p$ indicate elastic and plastic components. Both elastic and plastic flexure occur in response to the same total bending moment $M$.  The elastic curvature is related to $M$ according to \citep{Mansfield1989}
\begin{equation}
    \label{eq:elastic-constitive}
    \curvature_e = \curvature_n - \frac{M}{EI},
\end{equation}
where $E$ is a Young's modulus modified for plane strain, $I$ is the second moment of area, and $\curvature_n$ is the neutral curvature of the plate \citep{Turcotte2002}. As proposed by \cite{Buck2001}, the neutral curvature enables the plate to have zero elastic stress when it is born on the isostatic high of the ridge axis.  We treat $E$ as constant and uniform and, importantly, allow the second moment of area to vary in space and time according to 
\begin{equation}
    \label{eq:second-moment-area}
    I(x,t) = h^3/12.
\end{equation}
The second moment of area inherits a dependence on $x-Ut$ from $h$.

Viscoplastic curvature $\curvature_p$ accumulates in the material with progressive flow when and where $M$ exceeds the yield moment $\yield$; this is a Bingham-type plasticity. Hence we track $\curvature_p$ as it moves with the plate,
\begin{equation}
    \label{eq:viscoplastic-constitutive}
    \ld{\curvature_p}{t} = -\frac{M}{\eta I}F(M/\yield),
\end{equation}
where $F()$ is the yield function, $\eta$ is the viscosity above yield (a constant), and the Lagrangian derivative is defined for any quantity $a$ as
\begin{displaymath}
    \ld{a}{t} \equiv \pd{a}{t} + U\pd{a}{x} = \dot{a} + Ua'.
\end{displaymath}
Following \cite{Hewitt2020}, the yield function is
\begin{equation}
    \label{eq:yield-function}
    F(m) = \operatorname{max}\left(0,\,1-{|m|}^{-1}\right).
\end{equation}
To clarify the meaning of this equation, we consider the following cases. For moments $|M|<\yield$ we have $MF(M/\yield)=0$ and, by equation~\eqref{eq:viscoplastic-constitutive}, we have zero rate of viscoplastic flexure. In contrast, when $M>\yield$ we have
\begin{displaymath}
    M F(M/\yield) = M\left(1-|\yield/M|\right) = M - \yield.
\end{displaymath}
Similarly, when $M<-\yield$ we have $M F(M/\yield) = M + \yield$. Hence $MF(M/\yield)$ is a measure of bending moment in excess of the yield moment.

To obtain the combined rheological model relating $M$ to $\curvature$, we substitute \eqref{eq:elastic-constitive}, \eqref{eq:viscoplastic-constitutive} and \eqref{eq:yield-function} into the Lagrangian derivative of \eqref{eq:maxwell-total-curvature}. Noting that the Lagrangian derivatives of $\curvature_n(x,t)$ and $I(x,t)$ are zero, we obtain 
\begin{equation}
    \label{eq:elastic--viscoplastic-constitutive}
    \left[\frac{1}{E}\ld{}{t} + \frac{1}{\eta} F(M/\yield)
    \right] M = -I\ld{\curvature}{t}.
\end{equation}
This equation, representing the Maxwell combination of elastic and viscoplastic flexure, must be satisfied at all $t$ and all $x>0$.

The next step in developing our theory arises from the consideration that fractures and faults are self-localising. A fracture formed by an initially uniform stress field continues to extend at its tip, at the expense of any distributed deformation and other, proximal, incipient fractures. This is a consequence of the focusing of stress onto the fracture tip. To replicate this behaviour within our model, and hence to use plasticity to represent faulting, we introduce yield weakening.  This means that viscoplastic flexure at a point in the material causes a decrease in the yield moment $\yield$ at that point.  We model this strength degradation with
\begin{equation}
    \label{eq:yield-weakening}
    \ld{\yield}{t} = -\frac{\yield - \yield^\text{min}}{\weakcurvature}\left\vert\ld{\curvature_p}{t}\right\vert ,
\end{equation}
where $\yield^\text{min}>0$, a constant, is the yield moment associated with the limiting amount of yield weakening, and $\weakcurvature$ is a constant.

To illustrate the effect of the formulation \eqref{eq:yield-weakening} on evolution of the yield moment at a material point, we solve a simplified case.  Take a reference frame moving at  $U\xhat$ and assume that, for illustration, $\dot{\curvature}_p$ is monotonically increasing from zero.  Rearranging and integrating gives $\yield(\curvature_p) = \yield^\text{min} + [\yield(0)-\yield^\text{min}]\exp\left(-\curvature_p/\weakcurvature\right)$, where $\yield(0)$ is the yield moment when $\curvature_p=0$. This shows that yield weakening is exponential with progressive plastic flexure with an e-folding plastic curvature $\weakcurvature$, and that it saturates at the prescribed minimum yield moment.

In obtaining numerical solutions, the minimum value of the yield moment $\yield^\text{min}$ must be specified.  This minimum represents a physical limit on the self-localising effect of faults as they extend downwards. In this sense, $\yield^\text{min}$ might correspond to a fault depth where the lithostatic pressure or fault rotation inhibits further fault growth \citep{buck1993, lavier2000factors}. However, we avoid direct attribution and instead assume that the yield moment can decrease from its initial amount by a fraction $f_W$ of unity, such that $\yield^\text{min}(x,t) = (1-f_W) \yield(x,0).$

\subsection{Force balance and boundary conditions}
\label{sec:theory-forces}

Vertical force balance, excluding any net, in-plane stress, takes the simple form $M'' + q = 0,$ where $q$ is a distributed load on the plate in the positive $z$ direction \citep{Mansfield1989, Turcotte2002}. Since we assume that the plate and mantle beneath have the same density, the restoring force for vertical deflections is $-\Delta\rho g w$, where $\Delta\rho = \rho - \rho_o$ is the difference with the ocean density.  Variations in plate thickness $\pert{h}(x,t)$ away from the mean plate thickness $h_0$ contribute to the load insofar as they protrude from the top of the plate and hence displace water.  We assume that thickness perturbations are symmetric about the plate mid-plane, such that half contributes to surface elevation and hence to buoyancy forcing. This is a simplifying assumption consistent with classical beam theory that has a negligible effect on our results (Appendix~\ref{sec:linearised-analysis}). Then the vertical force balance becomes
\begin{equation}
    \label{eq:vertical-force-balance}
    M'' - \Delta\rho g \left(w + \pert{h}/2\right) = 0.
\end{equation}
Following equations \eqref{eq:elastic--viscoplastic-constitutive} and \eqref{eq:yield-weakening}, this is our third governing equation. 

This force-balance equation requires boundary conditions at both the ridge axis ($x=0$) and in the far field ($x\to\infty$).  The latter should enforce a regime in which the only variation is due to the uniform translation at speed $U$.  Hence the far-field condition is 
\begin{equation}
    \label{eq:bc-far-field}
    \ld{\,\cdot}{t} = 0 \qquad \text{for }x\to\infty,
\end{equation}
where $\cdot$ represents any appropriate quantity. In numerical solutions developed below, we impose this condition at a finite value of $x$.

Ridge-axis boundary conditions for this model were established by \cite{Buck2001}.  Partially molten mantle rock ascends to shallow depths at the ridge axis.  The low density $\rho_a$ of this hot, two-phase aggregate promotes a near-axis isostatic balance that differs from the balance attained far from the ridge axis. This makes the axis a topographic high, with relief $w(0,t)$ (above the far-field reference height). Isostatic balance gives
\begin{equation}
    \label{eq:bc-isostatic-height}
    w(0,t) = Rh(0,t),\qquad\text{with }R\equiv \frac{\rho-\rho_a}{\Delta\rho}.
\end{equation}
Here we have relaxed the assumption by \cite{Buck2001} that axial relief is constant. Instead we allow it to vary in time with the plate thickness of eqn.~\eqref{eq:plate-thickness-decomp}. 

As the plate moves away from the ridge axis, it descends from its axial high and settles into a far-field isostatic equilibrium at a lower height.  This transition occurs smoothly, along a flexural curve, like the curve of a stack of paper lifted off the table at one edge. \cite{Buck2001} assumes that the length-scale $\ell$ on which this bending occurs is roughly the flexural--isostatic scale,
\begin{equation}
    \label{eq:flexural-isostatic-length}
    \ell \equiv \left({EI}/{\Delta\rho g}\right)^{1/4}.
\end{equation}
In particular, he assumes that the curvature of the plate, where it forms at the ridge axis, is given as $\curvature_n \sim w(0)/\ell^2$.  This curvature, which \cite{Buck2001} refers to as accretional, is ingrown into the plate, meaning that it creates zero flexural stress. We adopt this hypothesis and allow it to inherit a time dependence from $h(x,t)$,
\begin{equation}
    \label{eq:bc-isostatic-curvature}
    \curvature(0,t) = \curvature_n(0,t) = \frac{Rh(0,t)}{\left[{EI(0,t)}/{\Delta\rho g}\right]^{1/2}}.
\end{equation}
This is our boundary condition on $\curvature$, and it implies that 
\begin{equation}
    \label{eq:bc-zero-moment}
    M(0,t) = 0,
\end{equation}
i.e., the bending moment is zero at the ridge, where new plate is formed.

We further assume that the plate is born with zero plastic curvature, and therefore it is born with its maximum yield moment,
\begin{equation}
    \label{eq:bc-initial-yield-moment}
    \yield(0,t) = \yield^\text{max}(0,t) = \ystress h^2(0,t)/6,
\end{equation}
where $\ystress$ is the fibre stress at yield of the plate. This formulation relating $\yield$ to $\ystress$ ensures that yield moment scales with $h$ such that the material strength is invariant \citep{stok2009}.

Equations \eqref{eq:bc-isostatic-height} and \eqref{eq:bc-isostatic-curvature}--\eqref{eq:bc-initial-yield-moment} are the four ridge-axis boundary conditions that we need below. Of these, \eqref{eq:bc-isostatic-height}, \eqref{eq:bc-isostatic-curvature} and \eqref{eq:bc-zero-moment} are identical to those of \cite{Buck2001}, who did not include plastic yielding.  

\subsection{Parameter values and scales for fast-spreading mid-ocean ridges}
\label{sec:theory-parameters}

An exact assignment of dimensional values to the problem is impossible in the present context. Reference values used in this work are listed in Table~\ref{tab:model-parameters}.  The half-spreading rate $U$ varies amongst ridges, but \cite{Huybers2022} consider fast-spreading ridges to be those with $U\ge3.8$~cm/yr.  Similarly, sea-level varies at a range of frequencies, but its power spectrum has peaks at Milankovitch periods of 41 and $\sim$100 ka.  Densities of sea water, mantle rock, and sub-ridge magmatic mush are reasonably well known, but will vary to some extent in time and space; likewise for the elastic stiffness of oceanic plates \citep{Turcotte2002}.  Our choices for these parameters are thus representative values.  We follow \cite{Buck2001} in estimating mean plate thickness $h_0$ using the flexural--isostatic length \eqref{eq:flexural-isostatic-length} inferred from mean bathymetric profiles for fast-spreading ridges \citep{Small1994}.  We obtain a value that is consistent with values of elastic thickness of young lithosphere \citep{Burov2011bs}.

\begin{table}[]
    \centering
    \begin{tabular}{lcccl}
        Quantity & symbol & value or range & units & comment \\ \hline
         Flexural--isostatic length scale & $L$ & 10--15 & km & \cite{Buck2001} \\
         Reference plate thickness & $h_0$ & 3--5 & km & \cite{Buck2001} \\
         Half-spreading rate & $U$ & 6 & cm-yr$^{-1}$ & fast$^*$ is $\ge4$~cm-yr$^{-1}$ \\
         Period of sea-level variation & $T$ & 41 or 100 & kyr & \\
         Density of plate and ambient mantle & $\rho$ & 3200 & kg-m$^{-3}$ & \\
         Density of ocean water & $\rho_w$ & 1000 & kg-m$^{-3}$ & \\
         Density of partially molten rock & $\rho_a$ & 2800 & kg-m$^{-3}$ & \\
         Modified Young's modulus & $E$ & 25 & GPa & \cite{heap2020towards-58b} \\
         Fibre stress at which plate yields & $\ystress$ & & Pa & $^\dag$, see sec.~\ref{sec:discussion} \\
         E-folding curvature of plastic weakening & $\weakcurvature$ & & m$^{-1}$ & $^\dag$, see sec.~\ref{sec:theory-parameters} \\
         Viscosity of flexure above yield stress & $\eta$ & & Pa-s & $^\dag$, see sec.~\ref{sec:theory-parameters} \\ \hline
         Frequency and wavenumber & $\omega$ & 75 or 30 & - & $2\pi L/UT$ \\
         Axial isostatic density ratio & $R$ & 0.2 & - & $(\rho-\rho_a)/(\rho-\rho_w)$ \\
         Deborah number & $\deb$ & $2\times(10^4$--$10^5)$ & - & $EL/\eta U$\\
         Reference yield moment & $\yieldref$ & 1 & - & $\ystress h_0^2L^2/6EI_0h_0R$\\
         Plastic weakening scale & $\curvatureref$ & 1 & - & $\weakcurvature L^2/h_0R$\\ 
         Plastic weakening fraction & $f_W$ & $10^{-2}$ & - & $0<f<1$ \\ \hline
    \end{tabular}
    \caption{Dimensional and dimensionless model parameters. Notes: $^*$ This is the definition for fast spreading adopted by \cite{Huybers2022}. $^\dag$ We assign this dimensional parameter implicitly via the non-dimensional parameter in which it appears. }
    \label{tab:model-parameters}
\end{table}

The simplicity of our model means that the physics of fracture and frictional sliding on faults are abstracted into effective quantities \citep[e.g.,][]{poliakov98, lavier2000factors}. Therefore, for the associated parameter values, geophysically correct values are unclear a priori. For example, our viscoplastic representation of faulting uses the viscosity to impose a timescale on plastic flexure, as inertia and friction do for dynamic faulting \citep{segall2010earthquake}.  However, here we are uninterested in dynamics on the timescale of brittle processes, so the viscosity may be considered a regularisation of the  plasticity.  Therefore it must be taken as sufficiently small to separate the timescale of the build-up of elastic stress ($L/U$) from the Maxwell time for viscoplastic adjustment ($\eta/E$).  In other words, we require that the Deborah number, which is a dimensionless ratio of these two timescales, be large compared to unity. However, the cost of choosing a larger Deborah number is smaller space- and time-steps of our numerical solver due to sharper localisation of plastic deformation. To balance these considerations, we take $\deb=2\times10^4$ or larger.

We use a similar approach to selecting values for the fibre stress of yielding $\ystress$, and the curvature-scale $\weakcurvature$ over which the yield moment drops by plastic weakening.  We apply physical reasoning to the dimensionless quantities that govern the equations, under the assumption that our effective model should capture the characteristic features of the mid-ocean ridge and its off-axis normal faulting. This means that the dimensionless $\yieldref$ and $\curvatureref$ should be $O(1)$ to ensure that the processes that they describe (yielding; weakening) do occur under the circumstances of the model.  The choice of $f_W$ is particularly important because for a uniform plate, it controls the spacing of plastic kinks (effectively: faults), as we show in sec.~\ref{sec:periodic-kinks}.  We therefore constrain $f_W$ by requiring a mean kink-spacing of $\sim$2~km in models where $h=h_0$.

The characteristic scales that are used to define non-dimensional quantities are given in equation~\eqref{eq:characteristic-scales} of Appendix~\ref{sec:rescaling}.  The key choices are $h_0$ as a scale for plate thickness, $L = \sqrt{2}\ell(h_0)$ as a characteristic scale for distances from the ridge axis, and $L/U$ as a characteristic scale for times.  Other characteristic scales are denoted with square brackets, i.e., $[w],\,[I],\,[M]\,\text{and}\,[\curvature].$  In the text and figures below, we use both dimensionless and dimensional symbols, disambiguating only where necessary.

We do not conduct an exhaustive parameter sensitivity study on the full, elastic--viscoplastic model. Instead we explore selected sensitivities in the Results section, below, and in Appendix~\ref{sec:sensitivity}.  An important point is that for $\ystress\to\infty$, the governing equations revert exactly to the elastic form used by \cite{Buck2001}, as shown in Appendix~\ref{sec:elastic-reversion}. 

\section{Results}
\label{sec:results}

The governing equations are placed into non-dimensional form in Appendix~\ref{sec:rescaling}. Results are obtained by numerical solution of these equations except in section~\ref{sec:results-linearised}, where they are obtained analytically after linearising. The linearised analysis and numerical methods are described in Appendices~\ref{sec:linearised-analysis} and \ref{sec:numerical}, respectively.  Numerical results shown below have a dimensionless grid spacing $\delta_x = 10^{-3}$ or smaller. The code is available in an online repository \citep{code-repo}. 

Below we examine solutions for a uniform plate thickness. We show that unbending can create regularly spaced viscoplastic kinks, which we interpret as faults.  We then consider sinusoidal perturbations to plate thickness. Using the linearised analysis, we show that tensile fibre stress is enhanced where plate thickness is reduced.  We further show that such stress enhancements can control the spacing of viscoplastic kink-faults, even for small plate-thickness perturbations.  Finally, we consider plate thickness perturbations that might arise from sea-level variation, and show that these can pace kink-faults at Milankovitch frequencies.

\subsection{Uniform plate and periodic viscoplastic kinks}
\label{sec:periodic-kinks}

\begin{figure}
    \centering
    \includegraphics[width=0.8\linewidth]{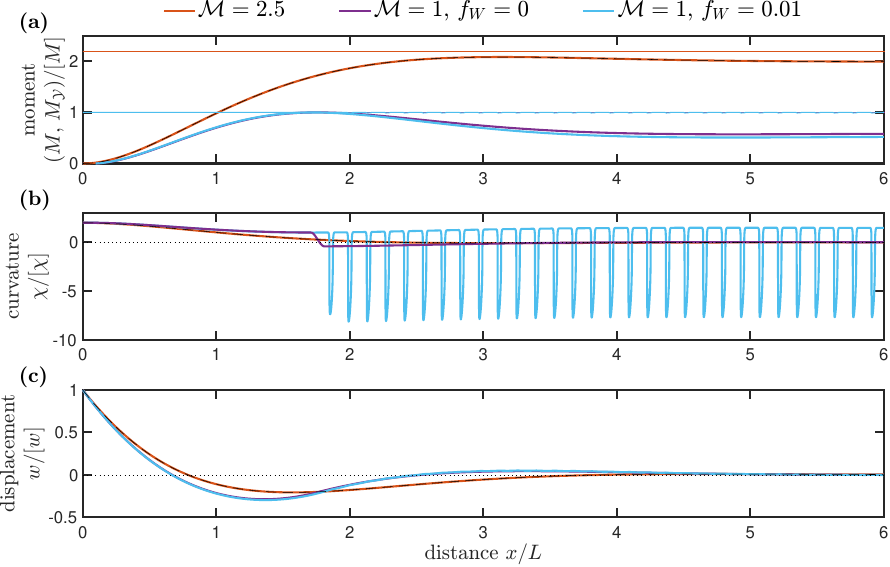}
    \caption{Solutions for a uniform plate, $h(x,t)=1$, at $t=16$ in three parametric cases. Case one has $\yieldref = 2.5$, which exceeds the maximum of $M$; case two has $\yieldref=1$ and no plastic weakening; case three has $\yieldref=1$ and 1\% plastic weakening. Analytical solutions for purely elastic unbending are shown as black dashed lines (which overlay red numerical curves). \textbf{(a)} Thick lines show the bending moment $M(x,t)$ for each case.  Thin lines show the (subtly evolving) yield moment $\yield(x,t)$.  Black dashed line is $M_0(x)=2\left[1-\e^{-x}\left(\cos x + \sin x\right)\right]$. \textbf{(b)} Curvature $\curvature(x,t)$. Black dashed line is $\curvature_0(x) = 2\e^{-x}\left(\cos x + \sin x\right).$ \textbf{(c)} Vertical displacement $w(x,t)$. Black dashed line is $w_0(x) = \e^{-x}\left(\cos x - \sin x\right).$}
    \label{fig:uniform-plate-cases}
\end{figure}

Representative solutions for three parametric cases are shown in Figure~\ref{fig:uniform-plate-cases}. All three cases have uniform plate thickness $h=1$. The first case, plotted as red curves, has an initial yield moment, $\yieldref = 2.5$, that exceeds the maximum moment achieved by elastic unbending of the plate.  Hence this case is purely elastic.  Its moment, curvature, and vertical displacement curves are independent of time and follow the analytical solution used by \cite{Buck2001}.

In the second case, plotted as purple curves, the initial yield moment is $\yieldref = 1$, and hence the moment $M$ readily reaches this threshold. This causes plastic flexure, evident in the curvature plotted in panel~(b) of Fig.~\ref{fig:uniform-plate-cases}. The step in curvature $\curvature$ at $x\approx1.7$ represents a plastic flattening of the plate.  It is independent of time as the plate moves through it.  This case has no plastic weakening (i.e., $f_W=0$) and hence there is a uniform distribution of plastic flexure beyond the step.

The third case introduces slight plastic weakening, allowing the yield stress to drop by 1\% over the first dimensionless unit of $\curvature_p$.  Plastic failure is thus promoted where it has previously occurred, which immediately creates localised plastic kinks in the plate.  These negative spikes of curvature, evident in the blue curve of Fig.~\ref{fig:uniform-plate-cases}b, grow at $x\approx1.7$ but are abandoned as they are advected toward greater distances.  Each spike in curvature corresponds to a kink in the plate, and to a small, sharp peak in the vertical displacement (the kinks are not visible in panel~(c)). However, averaging over a larger scale, the kinks lead to a flattening of the plate and a reduction of bending moment and elastic stress. This is identical, in effect, to normal faulting of an unbending plate \citep{Buck2001}, and hence we consider these kinks as representing normal faults.

\begin{figure}
    \centering
    \includegraphics[width=0.95\linewidth]{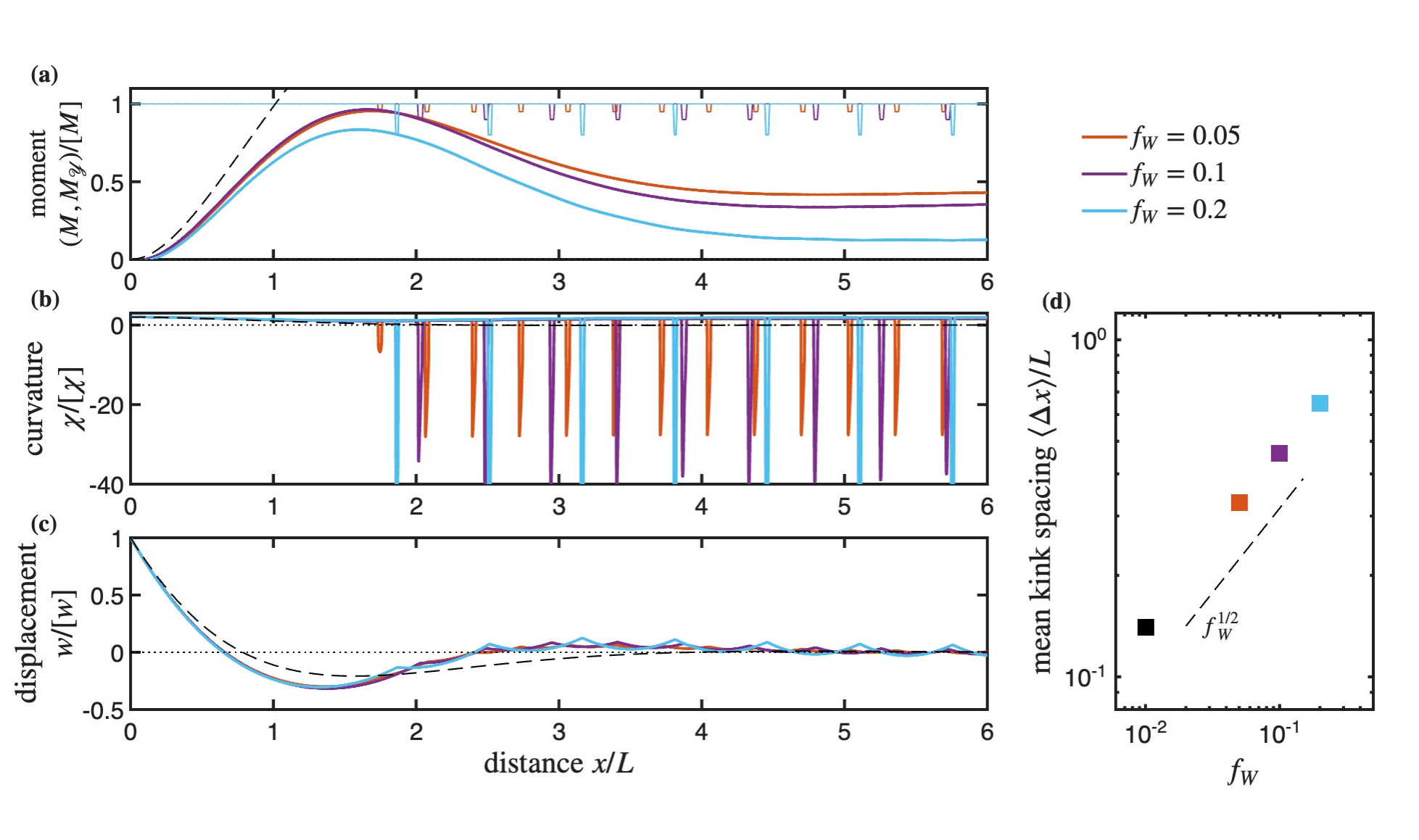}
    \caption{Solutions for a uniform plate, $h(x,t)=1$ at $t=14$ for three values of $f_W$. Other parameters as in Table~\ref{tab:model-parameters}. Dashed lines show the elastic solution. \textbf{(a)} Moment $M(x,t)$ and yield moment $\yield(x,t)$. \textbf{(b)} Curvature $\curvature(x,t)$. \textbf{(c)} Vertical displacement $w(x,t)$. \textbf{(d)} Mean spacing between viscoplastic kinks as a function of $f_W$. The black point corresponds to the reference case in Fig.~\ref{fig:uniform-plate-cases} with $f_W=0.01$. An animated version of this figure is provided in the online supplement as movie S1.}
    \label{fig:kink-spacing-fW-sensitivity}
\end{figure}

The flexural kinks become visible when yield weakening is larger.  Figure~\ref{fig:kink-spacing-fW-sensitivity} shows three solutions with $f_W=0.05,\,0.1,\,0.2$ that are otherwise identical to the third case of Fig.~\ref{fig:uniform-plate-cases}.  Panel~(b) shows that the negative spikes in $\curvature_p$ become larger and more widely spaced with increasing $f_W$.  Their topographic effect is visible in panel~(c). In panel~(d) the mean spacing of the kinks $\langle\Delta x\rangle$ is plotted against the weakening parameter; this indicates a power-law relationship $\langle\Delta x\rangle \propto f_W^{1/2}.$ We refer to this spacing, under uniform plate thickness, as the natural kink spacing. At present, we have no analysis that predicts this natural spacing.

A question that arises from these results is whether kink spacing remains uniform when plate thickness varies on a wavelength that differs from the natural spacing.  If plate-thickness variations can affect kink spacing---i.e., if kinks can phase-lock with plate thickness---what is the minimum amplitude of perturbations $|\pert{h}|$ that is required?  We address this in the next section.

\subsection{Monochromatic perturbations}
\label{sec:monochromatic}

Here we consider flexure of a plate with monochromatic perturbations to plate thickness.  In particular, we assume that the dimensionless plate thickness is the real part of 
\begin{equation}
    h(x,t) = 1 + \epsilon\e^{i\omega(x-t)},
\end{equation}
where $\epsilon$ is the amplitude of perturbations and $\omega \equiv 2\pi L/(UT)$ is a dimensionless angular frequency.  The dimensional, temporal period of oscillations in plate thickness is $T$, and this hypothetically corresponds to a period of sea-level oscillation (e.g., 41~ka). With the current scheme of non-dimensionalisation, $\omega$ is also the dimensionless wavenumber. For $L=12$~km and $U=5$~cm/yr, a period $T=41$~ka gives $\omega\approx37$ (note that this frequency is much larger than unity).

We analyse the consequences of this harmonic perturbation in two ways.  First, we suppress viscoplastic yielding and study the variations in plate fibre stress due solely to elastic unbending. This is achieved analytically by linearising the problem.  Then we restore yielding and consider the kink spacing as a function of $\epsilon$ and $\omega$. These calculations show that the stress response is linear in thickness perturbation, but the resulting kink-fault pattern is a consequence of nonlinear phase-locking.

\subsubsection{Linearised analysis of elastic flexure} \label{sec:results-linearised}

Full details of the linearised analysis are provided in Appendix~\ref{sec:linearised-analysis}. The basic idea is that we take $\epsilon\ll1$ and linearise the equations around the \cite{Buck2001} solution for unbending of a uniform plate. Importantly, we adopt the boundary conditions proposed by \cite{Buck2001}, and faithfully apply to these our perturbation and linearisation.  This means that the isostatic height of the ridge axis $w(0,t)$ and the neutral curvature $\curvature(0,t)$ vary sinusoidally in time at frequency $\omega$.

\begin{figure}
    \centering
    \includegraphics[width=\linewidth]{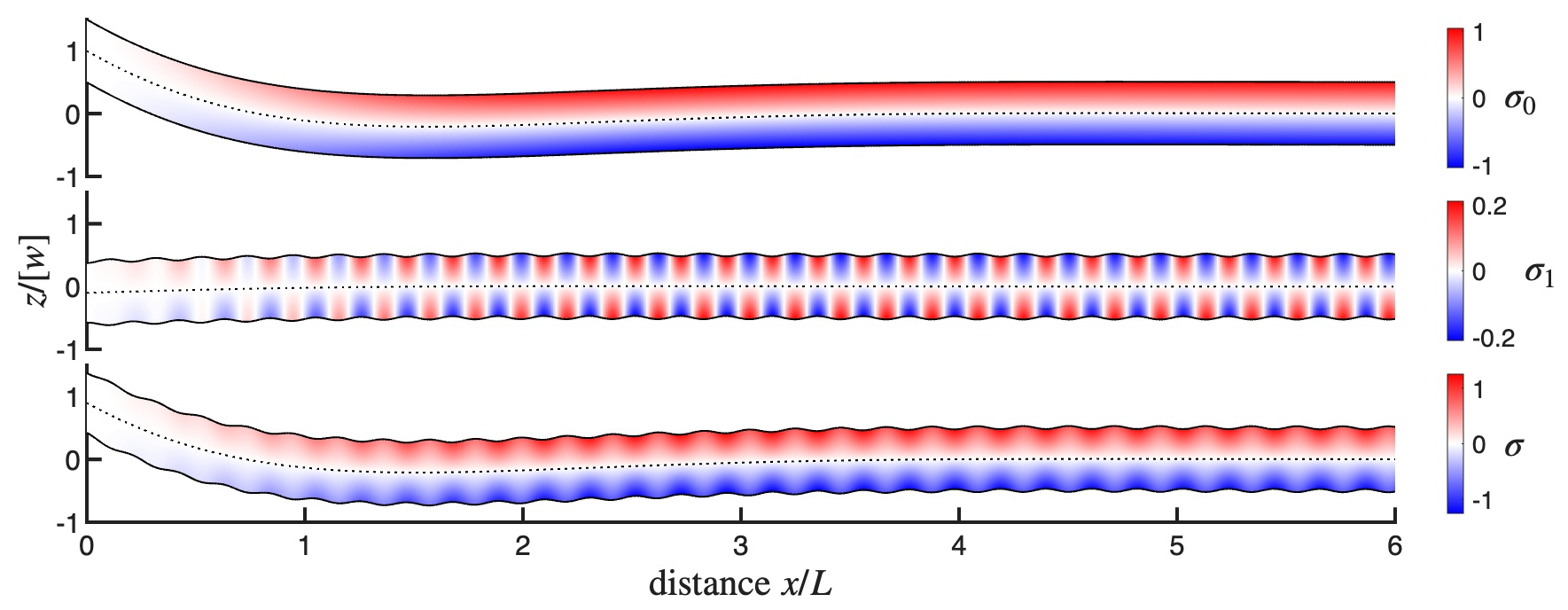}
    \caption{Fibre-stress maps from linearised analysis for a 10\% plate-thickness perturbation ($\epsilon=0.1$) with $\omega=30$ and $R=0.2$. Plate centreline is dotted. Red colour indicates tension; blue colour indicates compression. Top panel shows the base-state stress $\sigma_0$ in a plate displaced vertically by $w_0$. Middle panel shows the perturbation stress $\sigma_1$ with displacement $w_1$.  Bottom panel shows the total stress $\sigma=\sigma_0 + \sigma_1$ with displacement $w_0+w_1$. An animated version of this figure is provided in the online supplement as movie S2.}
    \label{fig:linearised-stress}
\end{figure}

We show in Appendix~\ref{sec:linearised-analysis} that these perturbations at the boundary have a negligible effect on the fibre stress $\sigma$ that gives rise to normal faulting, as does the load $q(x,t)$ associated with the perturbation. The important contribution to variations in $\sigma$ comes from the second moment of area, $I(x,t)$, which was introduced in equation~\eqref{eq:second-moment-area}. In thin-plate theory, the dimensionless fibre stress associated with small deflections is $\sigma(x,t) = zM/I$, where $z=0$ on the mid-plane of the plate, and 1/2 on its upper surface \citep{Mansfield1989}. This is a quantity of interest because tensile fibre stresses drive normal faulting.

Figure~\ref{fig:linearised-stress}a shows the fibre stress in the unperturbed state $\sigma_0$.  Red indicates tensile fibre stress.  This is the solution obtained by \cite{Buck2001} and represents our reference state.  The perturbation in stress $\sigma_1$ arising from a 10\% thickness perturbation with $\omega=30$ is shown in panel~(b). This perturbation is time dependent, as evident in the supplementary movie S2.  The total stress ($\sigma_0 + \sigma_1$) is shown in panel~(c).  The key feature of this total stress is the $\sim$20\% variation in tensile stress at the upper surface of the plate, which does not decay with distance from the ridge axis.  This variation has the potential to alter the natural kink spacing $\langle\Delta x\rangle$. 

Defining $\Sigma$ as the dimensionless fibre stress at the upper surface of the plate, we find that in the relevant limit of $\omega \gg 1$,
\begin{equation}
    \Sigma(x,t) \sim 1-2\epsilon\cos\left[\omega(x-t)\right] - \e^{-x}\bigg\{\cos x + \sin x + \epsilon\bigg[\sin x\cos \omega t - 2(\cos x + \sin x)\cos[\omega(x-t)]\bigg]\bigg\}. 
\end{equation}
The terms in curly brackets, multiplied by $\exp(-x)$, are the near-field terms, which become negligible for $x\gtrsim 2$.  The remaining terms are of greater importance here. The unit term is the tensile stress arising from complete unbending of a uniform-thickness plate. The next term is the perturbation of this stress generated by variations in the second moment of area, $I_1(x,t)$. This perturbation has an amplitude $2\epsilon$ that is independent of frequency.  It therefore has the potential to pace faulting over a wide range of frequencies.  We next show that it can indeed do this, even at much smaller $\epsilon$ than that used in Fig.~\ref{fig:linearised-stress}.

\subsubsection{Pacing of viscoplastic kink-faults}
\label{sec:results-monochrome-pacing}

Above we have shown that \textit{(i)} fibre stresses arising from unbending of a uniform plate can cause regularly spaced kink-faults (\S\ref{sec:periodic-kinks}), and \textit{(ii)} variations in plate thickness can induce variations in fibre stress of the same relative amplitude (\S\ref{sec:results-linearised}). From these, it follows that variations in plate thickness can potentially pace faulting.  Here we test the amplitude of thickness perturbation $\epsilon$ that is required to pace faulting.

\begin{figure}
    \centering
    \includegraphics[width=0.95\linewidth]{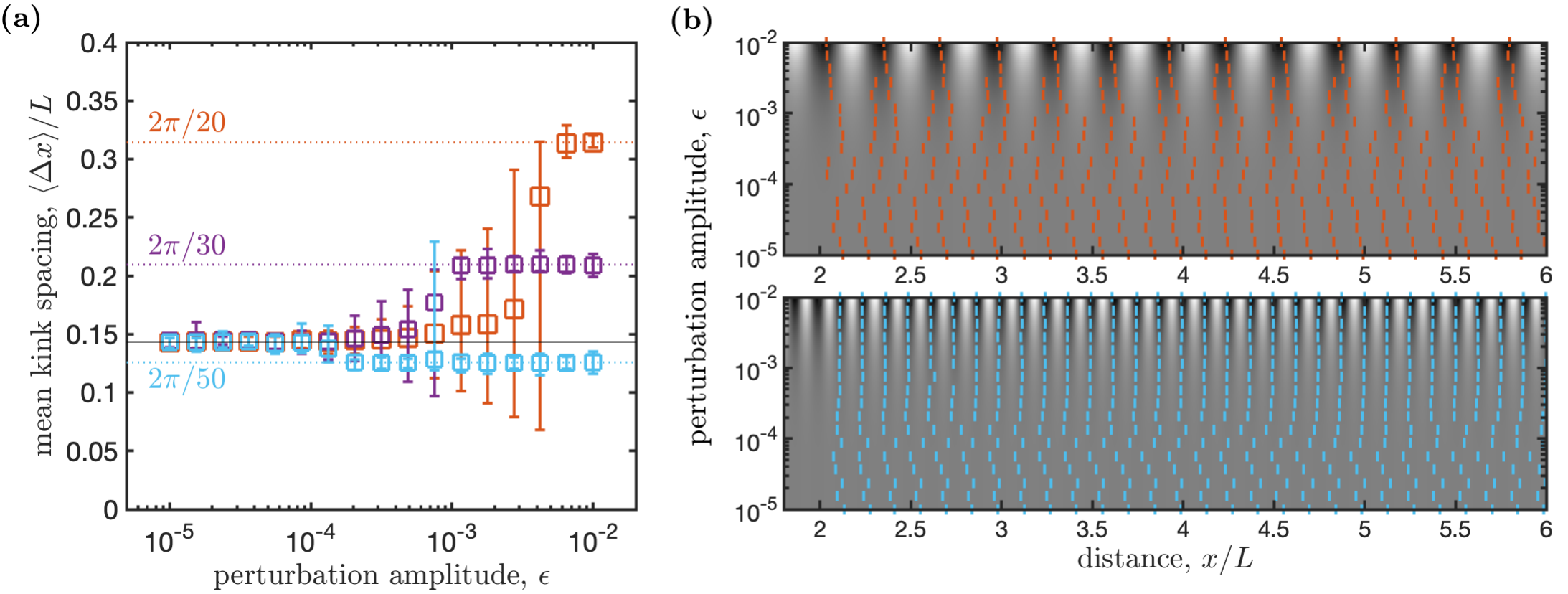}
    \caption{Kink phase-locking with plate-thickness variations at increasing perturbation amplitude $\epsilon$ with $\deb=2\times10^5$.  Results for three values of $\omega$ are shown in different colours: $\omega=20$ in red; $\omega=30$ in purple; $\omega=50$ in blue. Full domain width is 8 and time is 14. Other parameters as in Table~\ref{tab:model-parameters}. \textbf{(a)} Mean kink spacing $\Delta x$ as a function of $\epsilon$. Error bars indicate the minimum and maximum spacing. 
    Black line gives natural spacing without perturbations; coloured lines indicate spacing at the perturbation wavelength. \textbf{(b)} Kink positions for two values of $\omega$ (20 top; 50 bottom), plotted at each value of~$\epsilon$. Note that each row of points is computed independently. Background greyscale shows plate thickness variation, where black indicates negative (thin) perturbations.}
    \label{fig:monochromatic-amplitude-sensitivity}
\end{figure}

Figure~\ref{fig:monochromatic-amplitude-sensitivity}a plots the mean spacing of kink-faults as a function of $\epsilon$ for three values of dimensionless perturbation frequency ($=$wavenumber). When the perturbation amplitude is vanishingly small ($10^{-5}$), the spacing of kink-faults is well predicted by the spacing obtained for a uniform plate (black point in Fig.~\ref{fig:kink-spacing-fW-sensitivity}a). When the perturbation is relatively large ($10^{-2}$), the spacing of kink-faults is closely matched to the wavelength of perturbations. Figure~\ref{fig:monochromatic-amplitude-sensitivity}b shows that this spacing emerges because kink-faults occur where the plate is thinnest.  These locations are where the upper surface of the plate has its maximum tensile fibre stress, as shown by the linearised analysis (Fig.~\ref{fig:linearised-stress}).  

The transition from the natural kink-fault spacing at small $\epsilon$ to the perturbation-induced spacing at larger $\epsilon$ is a process of non-linear phase locking.  Fig.~\ref{fig:monochromatic-amplitude-sensitivity}a,b shows that the transition occurs over a range of $\epsilon$.  Considering the red series with $\omega=20$, the wavelength of perturbations is about 2$\times$ larger than the natural spacing.  With increasing $\epsilon$,  the density of faults must decrease to achieve this spacing. Kink-faults migrate toward thinner points of the plate; pairs approach each other, reducing the minimum spacing and increasing the maximum spacing.  At $\epsilon\approx10^{-2}$, these pairs are replaced by single kink-faults, spaced at the perturbation wavelength. 

For the case of $\omega=30$ (purple), the perturbation wavelength is about 1.5$\times$ the natural spacing. The adjustment in kink-fault spacing required for phase-locking is therefore more subtle than for $\omega=20$.  Phase-locking is complete when $\epsilon$ rises to about $10^{-3}$. 

For $\omega=50$ (blue), the perturbation wavelength is smaller than the natural spacing by about 10\%. Hence the density of faults must increase as $\epsilon$ increases.  To achieve this, one new kink-fault must appear for every $\sim$10 that preexist.  This occurs at $\epsilon\approx2\times10^{-4}$, when the kink-faults align to each thin point of the plate.

\begin{figure}
    \centering
    \includegraphics[width=\linewidth]{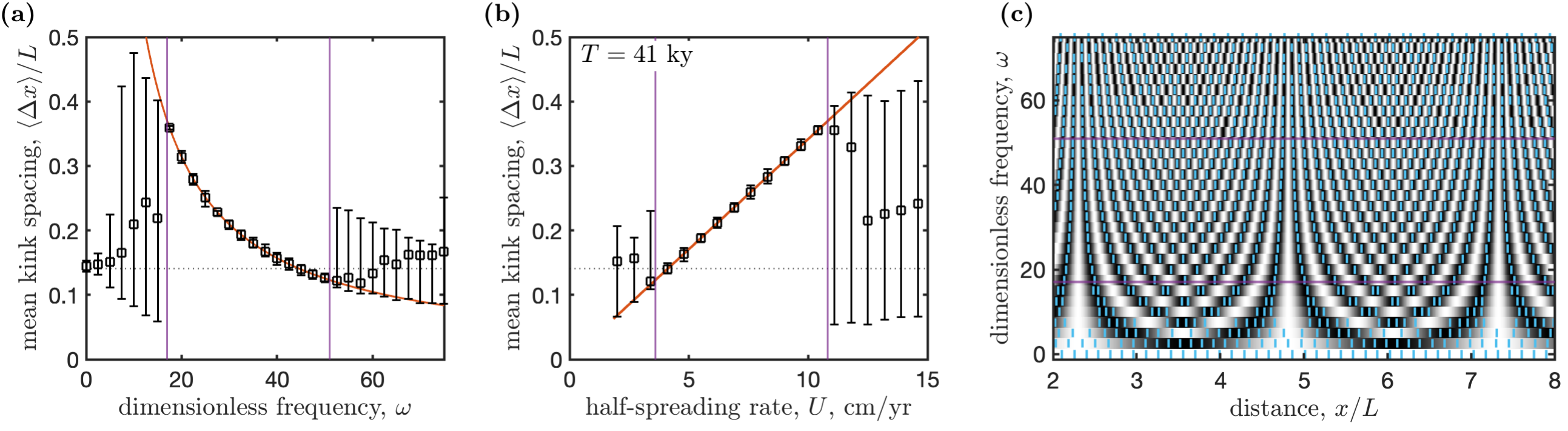}
    \caption{Kink-fault phase locking with monochromatic perturbation at fixed $\epsilon=0.01$, for a range of perturbation frequencies with $\deb=2\times10^5$. \textbf{(a)} Mean of kink spacing $\langle\Delta x\rangle$ as a function of perturbation frequency, which is equal to wavenumber. Error bars denote maximum and minimum spacing. Red curve plots dimensionless kink spacing of $2\pi/\omega$, the perturbation wavelength. Black dotted line is the natural spacing in the absence of perturbations. Purple lines bound the range in $\omega$ over which phase-locking is evident. \textbf{(b)} Symbols, lines, and $y$-axis as in panel~a, but now plotted in terms of dimensional half-spreading rate. This is calculated according to $U = 2\pi L/T\omega$, where $L=12$~km is the flexural--isostatic bending length and $T=41$~ky is the monochromatic period of forcing. \textbf{(c)} Kink positions plotted at $t=40$ for each value of~$\omega$ from panel~a. Each row of points is computed independently. Background greyscale shows plate thickness variation, where black indicates negative (thin) perturbations. Purple lines have the same meaning as in panels~a and b.}
    \label{fig:omega-phase-lock}
\end{figure}

In all three cases of $\omega$ shown in Figure~\ref{fig:monochromatic-amplitude-sensitivity}, phase-locking with the perturbation is complete when $\epsilon\gtrsim 10^{-2}$.  Using a perturbation amplitude $\epsilon=10^{-2}$, Figure~\ref{fig:omega-phase-lock}a shows the kink-spacing as a function of forcing frequency $\omega$.  The red curve represents the perturbation wavelength.  For frequencies in the range $\sim$18--51, phase-locking is evident.  At smaller $\omega$ (large wavelength), spacing varies widely and returns to the natural spacing as $\omega$ approaches zero.  At large $\omega$ (small wavelength), the minimum spacing tracks with the perturbation wavelength while the maximum is associated with multiples of this distance.  Fig.~\ref{fig:omega-phase-lock}c plots the position of kink-faults, calculated independently over a set of values of $\omega$, illustrating the relationship of spacing and plate thickness.

To confirm the relevance of this frequency range to the natural system, we consider a dimensional period $T=41$~ka in panel~b.  With the choice of flexural--isostatic bending length $L=12$~km, we can consider the dimensionless frequency $\omega = 2\pi L/T U$ for a range of half-spreading rates $U$. Figure~\ref{fig:omega-phase-lock}b plots mean kink-fault spacing as a function of $U$.  Phase locking occurs for $3.6<U<11$~cm/yr.  This is consistent with the range noted by \cite{Huybers2022}.

The perturbation amplitude required for phase-locking corresponds to 0.1--1\% of the mean plate thickness---a sensitivity that may be consistent with our hypothesis of sea-level-driven pacing of faults at fast-spreading MORs.  This sensitivity indicates that the system is responsive to small, coherent perturbations, rather than implying that such perturbations must dominate over all other sources of variability in natural settings.  Numerical solutions at lower and higher spatial resolutions $\delta_x$ (not shown) indicate that the phase-locking threshold is numerically robust.  

The linearised analysis shows that the amplitude of stress perturbations is independent of wavelength (in the high-frequency limit). However, the numerical solutions with viscoplasticity show that the resulting fault spacing depends on nonlinear phase-locking between the forcing wavelength and the kink spacing. In the next section, we investigate the consequences of phase locking when plate thickness varies with reconstructed Pleistocene sea level.

\subsection{Pleistocene sea-level perturbations}
\label{sec:results-sea-level}

Above we have shown that monochromatic perturbations in plate thickness $h$ can pace kink-fault formation in our models. We now turn to the question of whether thickness perturbations associated with Pleistocene sea-level (SL) variations might also pace kink-faults, and whether this might lead to kink frequencies aligned with the 41-ka Milankovitch band. Our first step is to use the $\delta^{18}$O record of Pleistocene sea level to construct a hypothesised plate-thickness variation.

\begin{figure}
    \centering
    \includegraphics[width=0.95\linewidth]{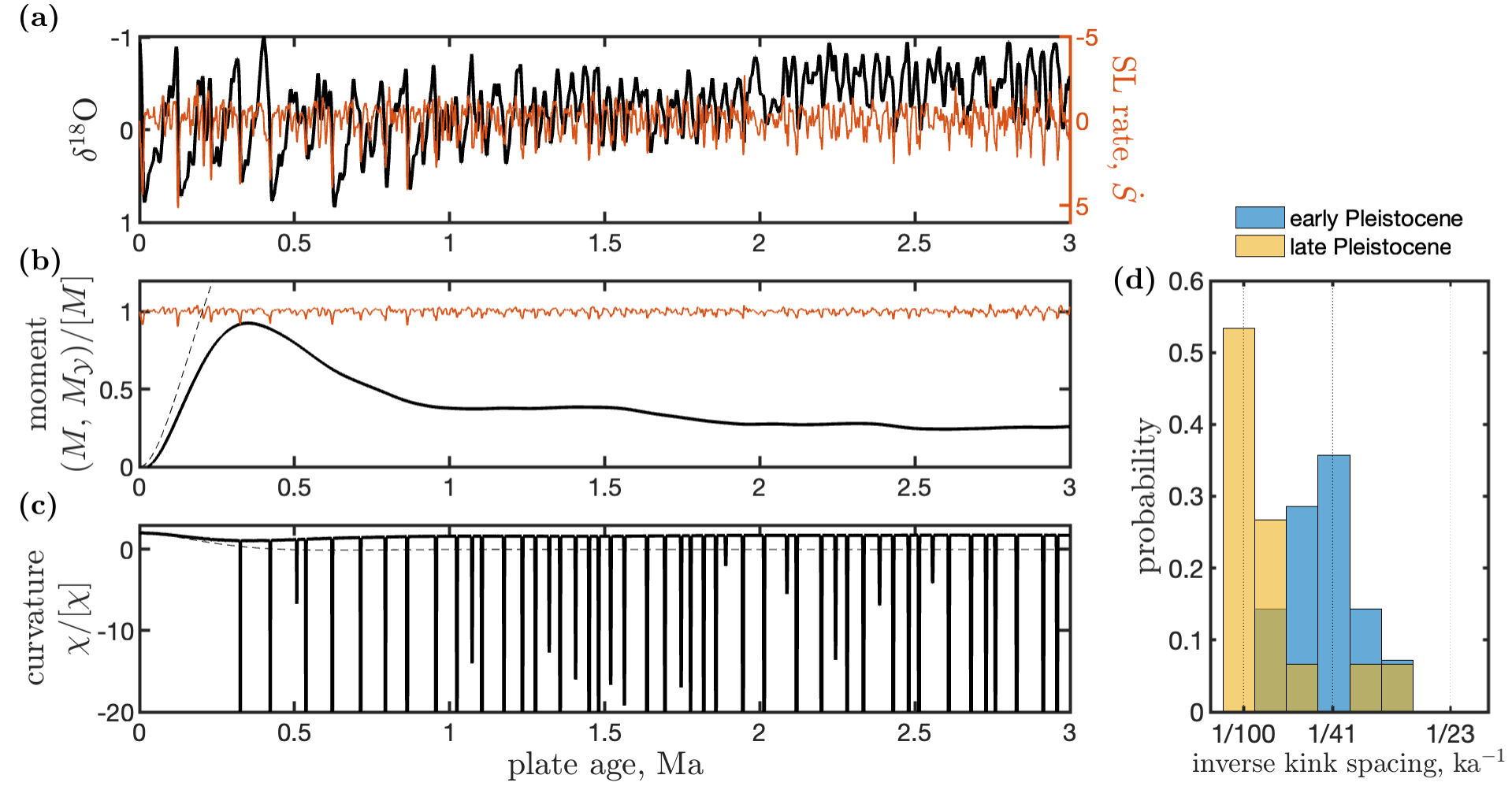}
    \caption{Plate-thickness perturbation derived from Pleistocene sea-level variation and its consequences in an unbending model with $\epsilon=0.01$.  Distance from the ridge axis is re-cast as dimensional age of the plate. Dashed curves are the elastic solution for a uniform plate.  \textbf{(a)} Black curve is the $\delta^{18}$O record from \cite{Huybers2007}; red curve is $\dot{S}$, the $Z$-score of sea-level rate.  The latter is obtained from $\delta^{18}$O by de-trending, extending (Appendix~\ref{sec:sea-level-extension}), time-differentiating, de-meaning, and rescaling by its standard deviation. \textbf{(b)} Yield moment $\yield(x,t)$ in red and the bending moment $M(x,t)$ in black. \textbf{(c)} Plate curvature $\curvature(x,t)$. \textbf{(d)} Histogram of kink-fault spacing $\Delta x$ in terms of its temporal frequency, separated into early (2.5--1.2~Ma) and late Pleistocene (1.2--0~Ma) periods. An animated version of this figure is provided in the online supplement as movie S3.}
    \label{fig:sealevel-flexure}
\end{figure}

Melting at mid-ocean ridges is driven by depressurisation of upwelling mantle.  Perturbations to this steady depressurisation can arise from sea-level variation on glacial--interglacial time-scales \citep[][and refs.~therein]{Katz2022}.  Melting-rate perturbations should then scale with (minus) the rate of sea-level change. To calculate this rate, we adopt the reconstruction of $\delta^{18}$O since 2.58~Ma by \cite{Huybers2007} as a proxy for global-mean sea level (Fig.~\ref{fig:sealevel-flexure}a, black curve), and de-trend it by subtracting off the best linear fit. We then extend the record using synthetic data---backwards in time to 10 million years ago and forwards in time to 7.5 million years in the future. The backwards extension has the spectrum and amplitude distribution of the early Pleistocene, and the forwards extension similarly mimics the late Pleistocene; details are in Appendix~\ref{sec:sea-level-extension}. With this extended record, we numerically differentiate, remove the mean, and normalise by the standard deviation.  We refer to this dimensionless, $Z$-score time-series as the SL rate, $\dot{S}$, and plot a segment in Fig.~\ref{fig:sealevel-flexure}a (orange curve). 

We now test whether the SL-rate, if transmitted to plate thickness, can organise fault spacing through the mechanics of unbending.  In panels (b) and (c) of Figure~\ref{fig:sealevel-flexure}, we plot results from an unbending model where plate-thickness perturbations are proportional to $-\dot{S}$,
\begin{equation}
    \label{eq:sea-level-plate-thickness}
    h(x,t) = 1 - \epsilon \dot{S}(x-t).
\end{equation}
The $x$ coordinate is now the dimensional crustal age, meaning the time since each material point was at the ridge axis.  The output time for this model is chosen to correspond to the present day, such that crustal age corresponds exactly to age in the sea-level reconstruction.  The orange curve in panel~(b) plots the yield moment $\yield(x,t)$; it inherits a variation from the square of the plate thickness (see eq.~\eqref{eq:bc-initial-yield-moment}) with $\epsilon=0.01$. The bending moment (black curve) can exceed the yield moment, but where and when it does, the excess moment drives viscoplastic flexure at a rate that is proportional to $\deb$.  A kink-fault emerges and the moment curve drops rapidly back below yield. Therefore, as with the monochromatic case (Fig.~\ref{fig:monochromatic-amplitude-sensitivity}), kink-faults form where the plate is thinnest (i.e., where the yield moment is smallest or, equivalently, where the fibre stress is largest).  

The kink-faults associated with this model are shown in panel~(c), and a histogram of their dimensional spacing is shown in panel~(d).  The spacing is plotted in terms of the inverse of the temporal separation between subsequent kink-faults, $(\Delta x/U)^{-1}$. On this axis, the fault spacing can be compared directly with the Milankovitch frequencies at which the $\delta^{18}$O record has power-spectral peaks.  We divide the set of $\Delta x$ values into those occurring in the early Pleistocene (ages 2.6--1.2~Ma) and the late Pleistocene (ages 1.2--0~Ma). The histogram shows a 41~ka peak for the early Pleistocene and a 100~ka peak for the late Pleistocene.  These peaks are consistent with the spectrum of the $\delta^{18}$O record---and with our hypothesis of sea-level pacing. 

\begin{figure}
    \centering
    \includegraphics[width=0.95\linewidth]{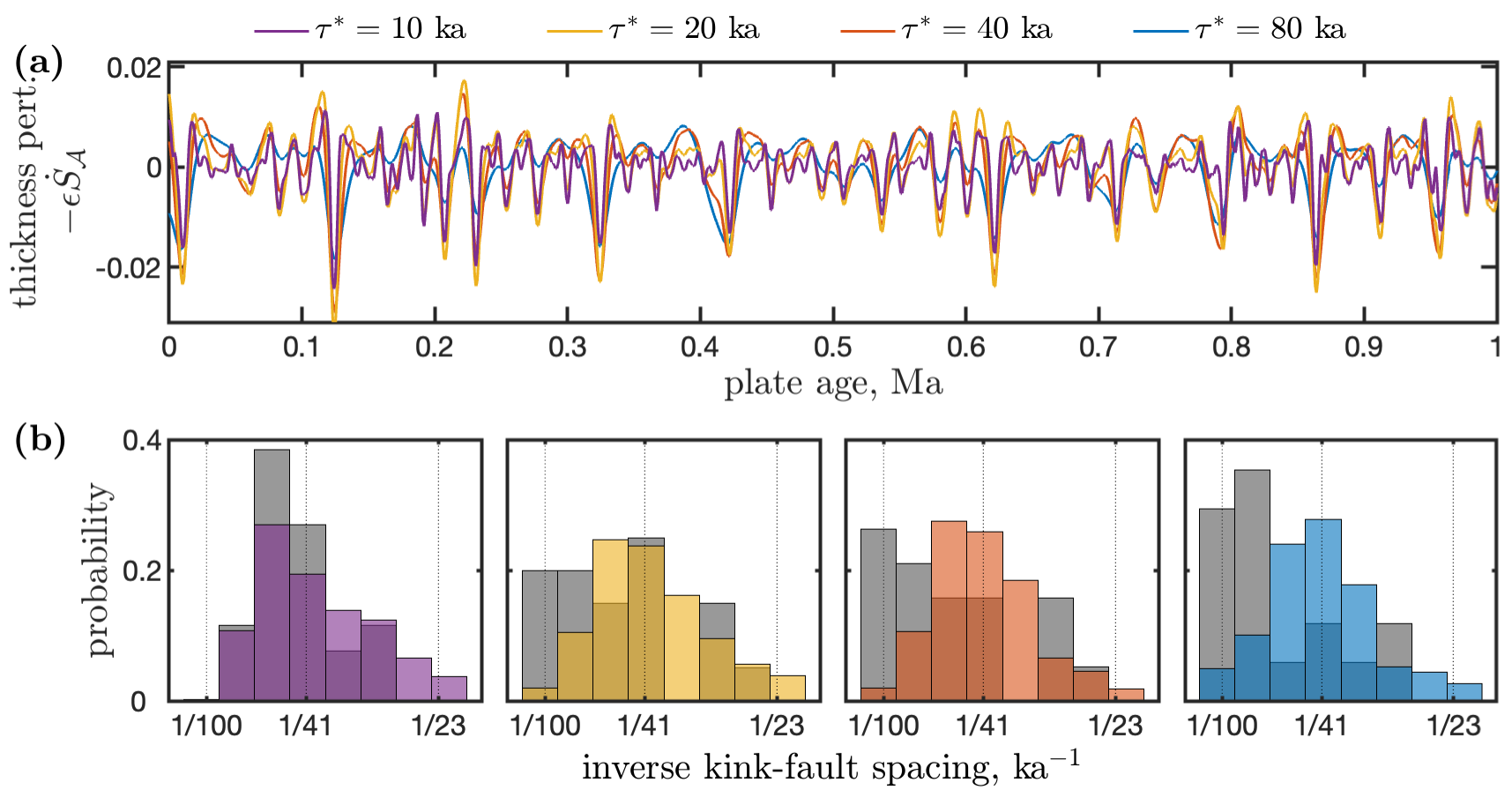}
    \caption{Sea-level forcing filtered by the melt-transport admittance spectrum of \cite{Cerpa2019} and \cite{Katz2022}, and its consequences for kink-fault spacing with $\epsilon=0.01$. Vertical dotted lines mark the Pleistocene frequencies of the eccentricity, obliquity and precession Milankovitch cycles. \textbf{(a)} Plate-thickness perturbation obtained by filtering $\dot{S}$ with the admittance spectrum \eqref{eq:admittance}. Colours correspond to four values of the peak admittance, $\omega^* = 2\pi/\tau^*$, for $\tau^*$ a characteristic melt-transport timescale. Only 1~Ma of the record is shown. \textbf{(b)} Histograms of kink-fault spacing in terms of their temporal frequency. Coloured bars count the full 17.5 Myr of the extended SL time-series (Appendix~\ref{sec:sea-level-extension}); grey bars count only the late Pleistocene (1.2--0~Ma).}
    \label{fig:sea-level-attenuation}
\end{figure}

In producing Figure~\ref{fig:sealevel-flexure}, we have assumed that plate-thickness perturbations are proportional to $\dot{S}$. However, this is inconsistent with theories of magma transport through the mantle  \citep[e.g.,][]{Cerpa2019}.  Perturbations in melting rate are delivered to the MOR axis by melt transport, where they become variations in magma supply. It is these supply variations that hypothetically perturb the plate thickness. But the spectral admittance from melting rate to MOR magma supply is not uniform. Rather, it is peaked at a period that roughly corresponds with the timescale $\tau^*$ of melt extraction across the MOR melting regime \citep{Cerpa2019}. Using standard values for mantle permeability, \cite{Katz2022} obtained an approximately log-normal admittance distribution with its peak at $\tau^*\approx38$~ka (their fig.~8a).  However, larger permeability estimates were obtained by  \cite{reesjones2020}, corresponding to shorter characteristic transport times.

We represent the combined effects of melt generation and transport using an admittance filter applied to $\dot{S}$. In particular, we multiply the discrete Fourier transform of $\dot{S}(t)$ by the admittance kernel 
\begin{equation}
    \label{eq:admittance}
    \mathcal{A}(\omega) = \exp\left[ - \left(\frac{\ln (\omega/\omega_*)}{\ln \omega_\sigma}\right)^2\right],
\end{equation}
where $\omega_*=2\pi/\tau^*$ is the angular frequency of peak admittance and $\omega_\sigma$ is a measure of the width of the distribution. The time-series with admitted frequencies $\dot{S}_\mathcal{A}(t)$ is then obtained by inverse Fourier transform of the product, and this is used to replace $\dot{S}$ in equation~\eqref{eq:sea-level-plate-thickness}.

The dimensionless, attenuated perturbation to plate thickness, $-\epsilon\dot{S}_\mathcal{A}$, is plotted in Figure~\ref{fig:sea-level-attenuation}a for four values of the characteristic melt-transport time $\tau^*$. The case of $\tau^*=10$~ka admits higher frequencies whereas the case of $\tau^*=80$~ka admits lower frequencies.  Each of these perturbations is used in an unbending model and, as in Fig.~\ref{fig:sealevel-flexure}, it paces the kink-faults that emerge.  
Histograms of kink-fault spacing $\Delta x$ are plotted in Figure~\ref{fig:sea-level-attenuation}b.  Each panel corresponds (by colour) to a value of $\tau^*$. The lighter bars count spacings of all kink-faults (using the full 17.5~Myr $\dot{S}$ time-series), whereas the darker bars indicate the spacing of late-Pleistocene kink-faults.  The key observation from these panels is that the peak is at a (temporal) spacing of $\sim$41~ka independent of $\tau^*$. This is consistent with expectation, given the dominance of this frequency in the forcing spectrum and its ability to phase-lock the fault spacing. A secondary observation regarding Fig.~\ref{fig:sea-level-attenuation}b is that the skew of the distribution of $\Delta x$ does depend on $\tau^*$. Smaller $\tau^*$ favours high-frequency kinks, with more counts at the 1/23~ka$^{-1}$ side; larger $\tau^*$ favours low-frequency kinks, skewing toward the 1/100~ka$^{-1}$ spacing.  This trend is especially evident in the distribution for the late-Pleistocene. The $\delta^{18}$O record for this period has more power in the 100~ka band.

\begin{figure}
    \centering
    \includegraphics[width=\linewidth]{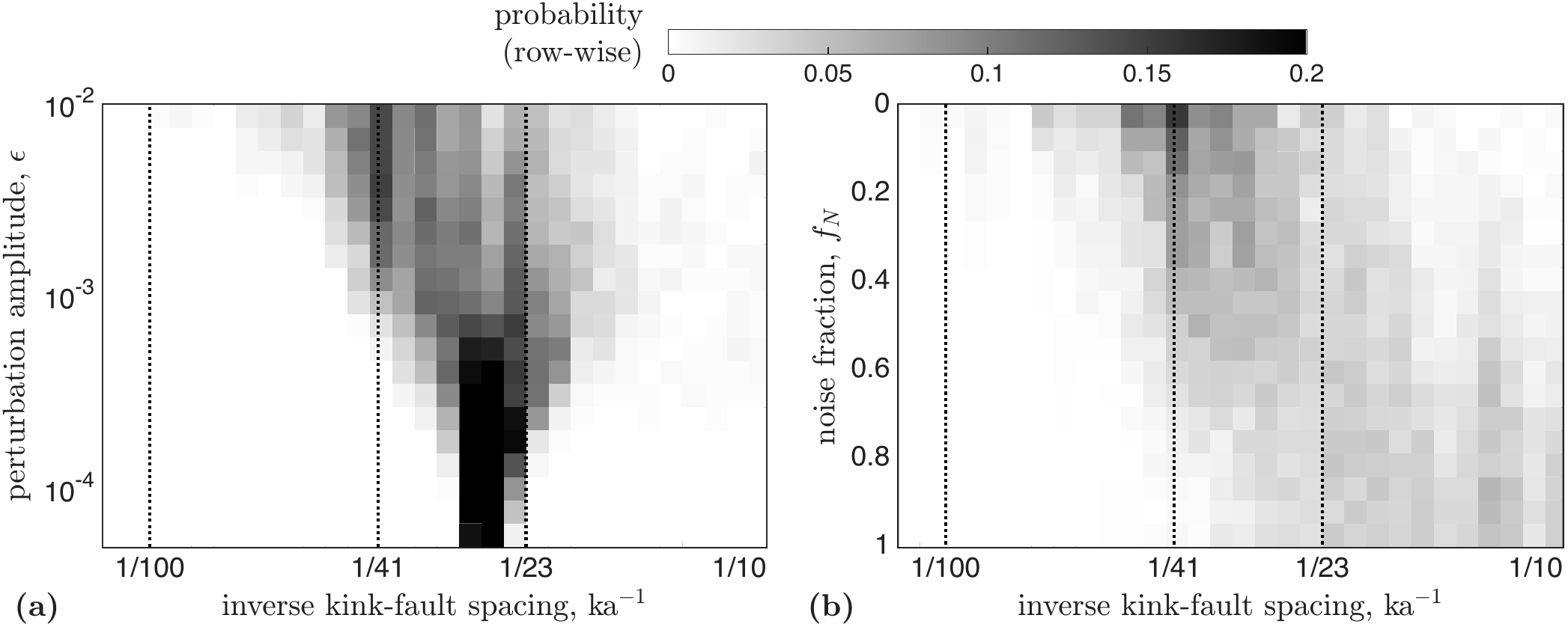}
    \caption{Sensitivity of kink-fault spacing for models forced by Pleistocene sea level. Plate thickness is generated via \eqref{eq:plate-thickness-noise} with an admittance-filtered sea-level-rate signal $\dot{S}_\mathcal{A}$ at $\tau^*=40$~ka (using the extended record from 6~Ma for 10~Myr). In both panels, rows are histograms approximating the probability distribution of kink-fault spacing (in terms of inverse kink-fault spacing in time), with probability coded by shading. Each row is derived from a single model; the suite of models varies according to the control parameter given on the $y$-axis. Vertical dotted lines mark the Pleistocene frequencies of the eccentricity, obliquity and precession Milankovitch cycles. In all cases $\deb=2\times10^4$. \textbf{(a)} Models run at different perturbation amplitudes, $\epsilon$, with zero additive noise.  The colour scale is saturated at the smallest perturbation amplitudes, where kink spacing is regular. \textbf{(b)} Models run with relative noise levels $0\le f_N\le 1$ and $\epsilon=0.01$.}
    \label{fig:noise-versus-sl}
\end{figure}

The Pleistocene SL record (and the admittance-filtered plate thickness) has a broadband spectrum.  We have shown above that for monochromatic forcing at $\tau \approx 41$~ky (Figs.~\ref{fig:monochromatic-amplitude-sensitivity}, \ref{fig:omega-phase-lock}), an amplitude $\epsilon \sim 10^{-3}$ is adequate for kink-fault phase-locking with the forcing signal.  To determine the minimum amplitude under the broadband SL forcing, we analyse models with $\epsilon$ ranging from $5\times10^{-5}$ to $10^{-2}$ in terms of their histograms of kink-fault spacing.  These are plotted in panel~(a) of Figure~\ref{fig:noise-versus-sl}. As observed for monochromatic forcing, the transition from the unforced spacing to phase-locking with plate thickness occurs at $\epsilon\sim10^{-3}$. 

The Pleistocene sea-level reconstruction data likely includes some high-frequency noise, and taking its time derivative enhances this noise relative to the signal.  Furthermore, the hypothesised transfer of $\dot{S}$ into plate thickness $h$ likely also introduces noise. To examine the sensitivity of fault spacing to this secondary noise, we construct a random time-series $N(t)$ with the same number of entries as the sea-level time-series. The noise series is random numbers distributed with uniform probability in the range $[-\sqrt{3},\, \sqrt{3}]$. This distribution has zero mean and unit standard deviation. The noise series $N$ is combined in a weighted sum with a filtered sea-level series $\dot{S}_\mathcal{A}$ to generate $h$ according to 
\begin{equation}
    \label{eq:plate-thickness-noise}
    h(x,t) = 1 - \epsilon\left[(1-f_N)\dot{S}_\mathcal{A} + f_N N\right].
\end{equation}
This noisy plate thickness is then input into the unbending model.

Figure~\ref{fig:noise-versus-sl}b shows row-wise histograms of kink-fault spacing for $f_N$ ranging from 0 (zero noise) to 1 (zero sea-level signal).  With $f_N=0$ in the top row, the histogram is peaked at a kink-fault spacing corresponding to $\sim$41~ka, just as in panel~(a) (but with different histogram bins).  In the bottom row, for $f_N=1$ (zero SL signal), the histogram is broad and shifted to higher frequencies, with no probability of a $\sim$41~ka spacing.  Increasing in noise content from $f_N=0$, the $\sim$41~ka peak diminishes, shifts to higher frequency, then vanishes. However, even when the noise level is quite high ($f_N \approx 0.2$), a kink-fault spacing of $\sim$41~ka still is most probable.  These results indicate that phase-locking persists in the presence of substantial stochastic variability, provided that the forcing retains modest coherence.

\begin{figure}
    \centering
    \includegraphics[width=0.95\linewidth]{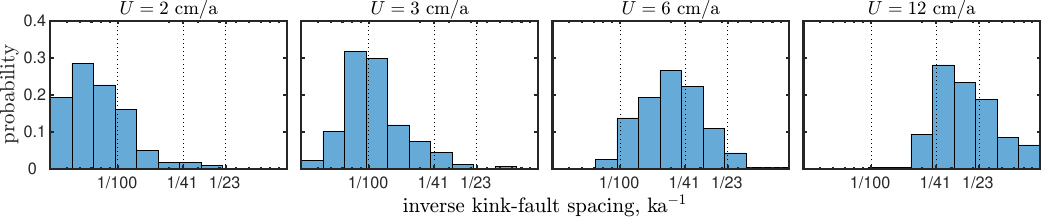}
    \caption{Histograms of kink-fault spacing $\Delta x$ for four  values of half-spreading rate $U$.  In each case, forcing is by the unfiltered, extended, $\dot{S}$ time-series applied as in \eqref{eq:sea-level-plate-thickness} with an amplitude $\epsilon=0.01$. What differs between plots is relationship $[t]=L/U$ between dimensionless model time and the dimensional time of sea level in the past. At faster spreading rate, the sea-level-driven perturbation is stretched over a proportionally larger distance away from the MOR axis, affecting which thickness deviations localise kink-faults. Note that the Deborah number $\deb$ is held constant at $2\times10^5$ across these models, despite its dependence on $U$. }
    \label{fig:spreading-rate-histograms}
\end{figure}

The results shown in Figure~\ref{fig:noise-versus-sl} indicate a robustness of kink-fault spacing to sea-level forcing: our results hold even for a small amplitude of plate-thickness perturbation and with substantial additive noise.  Those shown in Figure~\ref{fig:sea-level-attenuation} indicate an insensitivity to uncertainty (or variation) in the melt-transport timescale $\tau^*$.  However, Figure~\ref{fig:spreading-rate-histograms} shows that kink-fault spacing is sensitive to the assumed half-spreading rate $U$. At faster spreading rates, the distribution of kink-fault spacing shifts toward higher frequency (i.e., smaller temporal spacing).  This is qualitatively consistent with \cite{Huybers2022} (their Fig.~2), which found a dominant 41-ka spacing for fast-spreading ridges and a 100-ka spacing for intermediate spreading.  However, we must be cautious in interpreting Fig.~\ref{fig:spreading-rate-histograms} because the present model is based on the assumption of fast-spreading; it does not capture the effects of reduced $U$ on melt supply, ridge-axis morphology, or the mechanics and location of faulting \citep{Buck2005eo}.

\section{Discussion}
\label{sec:discussion}

Motivated by the significant concentration of spectral energy at the 41~ka period in bathymetric transects of fast-spreading ridges \citep{Huybers2022}, this manuscript develops a quantitative theory for a novel, mechanistic explanation. To assess the validity of this theory as a model of the natural system, we break the hypothesised mechanism into four pieces and refer to these in our discussion below. The pieces are: \textit{(a)} that SL-driven variations in melt-supply rate to the mid-ocean ridge cause variations in plate thickness; \textit{(b)} that as the plate moves away from the ridge axis (at $U>3.8$~cm/a) and unbends, these variations in plate thickness modulate elastic fibre stresses; \textit{(c)} that the variations in fibre stress pace faulting and hence control fault spacing; \textit{(d)} that fault spacing therefore inherits the signature of Pleistocene glacial--interglacial cycles. To test whether this hypothesised mechanism is physically plausible and adequate to explain the observations, we extend the unbending theory proposed by \cite{Buck2001} in two key ways. First, we extend the rheological model to include a viscoplastic mechanism designed to mimic faulting, and second, we impose a perturbation on the thickness of the plate such as might arise from variations in melt supply.  The results presented in the previous section illustrate the consequences of these modifications. We begin our discussion by considering the extent to which the results support the plausibility and adequacy of the hypothesis. We consider parts of the hypothesis in reverse order.

Regarding \textit{(d)}, the results clearly demonstrate that the spacing of kink-faults inherits the fingerprint of Pleistocene sea-level variations at a 41-ka period. This arises, in part, because of the prominence of the 41~ka peak in sea-level variation---a peak that is further accentuated by the admittance filter applied in forming the plate-thickness perturbation.  But it was not obvious, a priori, that this perturbation would be a control on the spacing of kink-faults. Under our proposed mechanism, it is a control because it transmits and organises components of the forcing spectrum through nonlinear phase-locking with the natural fault spacing (not because it selectively amplifies specific frequencies). Regarding \textit{(c)}, our results demonstrate that the position and spacing of kink-faults are sensitive to the variations in fibre stress.  These variations control the locations of the plate that first reach yield.  The consequent viscous kinking of the plate reduces the fibre stress in a neighbourhood around the kink. But ongoing plate motion (and elastic unbending) pushes the system back toward yielding, and fibre-stress variations again determine the position where yielding occurs. Above we show that for a monochromatic $T=41$~ky forcing (and our reference parameter values), phase-locking occurs for $U\gtrsim3.6$~cm/y, consistent with observations.  Regarding \textit{(b)}, our  linearised analysis elucidates the relationship between variations of plate thickness and fibre stress. This relationship is mediated by the second moment of area, $I\propto h^3$ (eq.~\eqref{eq:second-moment-area}). In Euler-Bernoulli plate theory, the fibre stress $\sigma$ scales with $M/I$, so reductions in $I$ translate to increases in $|\sigma|$. Details of the analysis demonstrate that this effect is independent of the wavelength of perturbation. A theory for wavelength selection of kink-faults must therefore invoke higher-order effects---a topic for future work.

Our discussion hence supports hypothesis parts \textit{(d)}, \textit{(c)} and \textit{(b)}, but only insofar as the mechanics of flexural, viscoplastic kinks are a good representation of the mechanics of faults.  This raises the first of two critical questions in assessing the validity of our model: To what extent are viscoplastic kinks a representation of sea-floor normal faults in the context of plate flexure?  A logical argument to address this question follows.  Fracture, faulting, and earthquakes are a thresholding phenomenon: as stresses rise, the rock is intact or the fault is locked until suddenly, it isn't. Then, once a fracture or slip-event initiates, it progresses irreversibly until a finite amount of work is done and stresses have dropped below the threshold by a finite amount. Two aspects of our theory account for this physics. First, the viscoplasticity introduces a threshold stress, cast as a yield moment. Above this yield moment, the deformation is viscous and hence dissipative and irreversible.  Second, the yield weakening means that points which have yielded to viscous flexure become more susceptible to further viscous flexure. Hence the bending moment undergoes a finite drop when yield is reached. This yield weakening leads to localisation, a key feature of faults whereby their sustained weakness promotes extension of the fracture and further slip---deformation which might otherwise have occurred elsewhere.  

These arguments for the physical similarity of viscoplastic flexural kinks and natural normal faults in oceanic crust are qualitative. A quantitative assessment is more difficult, and we shall not undertake a detailed consideration here. However, we can make a comparison of the parameter values implied in our models and those that might be chosen based on materials data. Following \cite{Buck2001}, we use the length-scale $L\approx 12$~km to match the topographic profile of fast-spreading mid-ocean ridges \citep{Small1994}. With this and other parameters as in Table~\ref{tab:model-parameters} we obtain $h_0\approx 4$~km, which is consistent with estimates of elastic thickness for young lithosphere \citep{Burov2011bs}. We can estimate the yield stress using this value of $h_0$ and other parameters, including $\yieldref=1$; we obtain $\ystress \approx 300$~MPa. Estimating the fault frictional resistance at a pressure corresponding to 4~km of water and 1~km of basalt gives a strength of about 60~MPa \citep{Byerlee1978}, neglecting fault orientation, fracture toughness, specific rock type, and other factors.  This differs from $\ystress$ by a factor of about five. Moreover, $\ystress$ is about a factor of three off of the yield-strength required in geodynamic models of plate tectonics \citep{Tackley2000wt}. We take this agreement to be semi-quantitative support for the physical similarity of viscoplastic kinks and oceanic normal faults.

The above discussion says nothing about part \textit{(a)} of the hypothesis, in which we \textit{assume} a relationship between MOR magma supply and plate thickness. This raises the second question of critical importance to the validity of our theory: how could plate thickness be modulated by magma-supply variations?  Before addressing this question, we should clarify the meaning of plate thickness in this context of fast-spreading MORs. Our theory adopts the idea from \cite{Buck2001}, which uses the word ``lithosphere'' to mean the portion of the plate that is elastic at the timescale of unbending, of order a few million years. \cite{Buck2001} argues that ``the cooling and accretion of lithosphere takes place over a zone of finite width [that is] short compared to the flexural wavelength of the lithosphere.'' In particular, the transition from the zone of ridge-axis isostasy and dike intrusion to elastic-plate flexure is assumed to occur within the first two kilometres distance from the axis.  The solidification and cooling of dikes across this transition is driven by intense hydrothermal circulation \cite[e.g.,][]{stein1994, maclennan2005}. Beyond this zone, slow cooling leads to a thickening of the elastic layer, but this is gradual enough that \cite{Buck2001} neglected it over the bending length-scale and treated the elastic plate thickness as a constant.

So if the plate in our model represents a dike-laden, hydrothermally cooled, dominantly elastic layer of rock, can it have variations in thickness?  Earth is heterogeneous at effectively all scales from grains to tectonic plates, so it seems safe to assume that the elastic thickness of fast-spreading oceanic lithosphere varies spatially. Moreover, \cite{Boulahanis2020gm} provided clear evidence for variation in crustal thickness---at the 10\% level---but crustal thickness and elastic thickness are distinct. Our analysis has shown that variations of the latter as small as 0.1\% are adequate to pace faulting.  Beyond this minimum amplitude, the plate-thickness variations must be coherent in the direction parallel to the ridge axis. Heterogeneity that is not coherent along ridge strike (e.g., due to magmatic solitary waves \citep{sim2022}) may lead to meander of fault traces, but would not pace faults on average.  Along-strike coherence of melt-supply variations is logically expected of sea-level forcing and, arguably, observed by \cite{Boulahanis2020gm}. 

One way that this might affect plate thickness is if the rate of axial dike injection varies with melt supply, and that such dikes promote small-scale fracture and enhanced permeability \citep{rubin1992, carbotte2006}.  Then the time-averaged hydrothermal circulation would be promoted in proportion to the rate of dike injection, and hence cool to a greater depth.  Alternatively, variations in magma supply might modulate the injection of heat without changing the time-averaged hydrothermal intensity, leading to variations in the extent of hydrothermal cooling. While this is facile speculation, it is less simple to envision a scenario in which magma supply varies \textit{without} leaving some trace in the thermal structure of the ridge and plate. To go beyond speculation and convincingly explore the physical details of such arguments is challenging and outside the scope of the current work.  However, we emphasise that the proposed mechanism does not require that sea-level forcing dominates all sources of variability, only that it provides a coherent perturbation capable of organising fault spacing through nonlinear dynamics. The present model assumes the coherence of such perturbations; departures from this assumption would likely reduce the efficiency of phase-locking but not the underlying mechanism. 

Let us now leave aside the question of SL-driven plate-thickness variation and suspend any lingering disbelief that our model formulation can capture the behaviour of flexural normal faults. Is it plausible that the 41~ka spectral concentration inferred for fast-spreading ridges reflects the natural pacing of faults in this context, instead of being evidence of SL control? Over the range of half-spreading rates from $\sim$4--8~cm/a, the characteristic spacing of faults varies linearly from $\sim$1--3~km, with a slope of $\sim$41~ka \citep{Goff2015, Crowley2015hf, Huybers2022}. There is nothing in the model of \cite{Buck2001} to predict this scaling, but \cite{Huybers2022} argued that an on/off magma supply controls fault timing at the ridge axis.  In the present analysis, the distance between kink-faults in a uniform plate is sensitive only to $f_W$ and $\yieldref$ (Fig.~\ref{fig:sensitivity}), neither of which are expected to vary with $U$ in the fast-spreading regime. We show in  Appendix~\ref{sec:sensitivity} that for SL- or noise-driven plate thickness variation, the kink spacing has a sensitivity to $\deb$ such that temporal spacing is $\propto -\log U$---but this scaling is not consistent with observations. We furthermore showed in Fig.~\ref{fig:spreading-rate-histograms} that $\sim$41-ka perturbations are most likely to be expressed in kink spacing despite small perturbation amplitude and substantial additive noise. Despite these lines of evidence, we cannot rule out the possibility that the observed 41-ka scaling emerges by an unknown mechanism that is distinct from the one proposed here.

Various avenues of further research could help to test the hypothesis of SL control of abyssal hills. Most relevant is improved observational coverage and resolution of bathymetric surveys, combined with detailed temporal analysis. In particular, insight could be gained by investigating the time-progression of normal faulting, both at the ridge axis and 30~km off-axis, where unbending drives fault slip \citep{crowder2000, escartin2007}. Existing or new seismic reflection profiles should be analysed for across-axis variations in elastic structure that are coherent along the axis \citep{Boulahanis2020gm}. 

Future theoretical work could better analyse the behaviour of the current formulation, particularly in terms of the preferred kink spacing and non-linear screening. It might also be fruitful to elaborate the current model.  The formulation of $h(x,t)$ could be enriched to include a slow thickening of the plate with distance from the ridge axis, along similar lines to \cite{Shah2003}. Another valuable refinement would be to better relate the yield moment and its weakening parameters $f_W$ and $\curvature_W$ to theories of oceanic normal-fault mechanics \citep{buck1993, Buck2005eo}.  Finally, we might hope for theoretical insight into how variations in magma supply translate (or not) to variations in plate elastic thickness via axial diking and hydrothermal cooling.

\section{Conclusion}
\label{sec:conclusion}

We have proposed and analysed a mechanism by which variations in plate thickness can modulate flexural stresses and thereby influence the spacing of normal faults at fast-spreading mid-ocean ridges. In this mechanism, unbending of the plate focuses tensile fibre stress in regions of reduced thickness, promoting localised failure. Using an elastic--viscoplastic flexural theory with yield weakening, we represent this failure as the formation of discrete kinks that serve as a proxy for faults.

Our results show that \textit{(i)} small-amplitude perturbations in plate thickness generate proportional variations in fibre stress, and \textit{(ii)} these stress variations can control the spacing of faults through nonlinear phase-locking. For perturbations of order $10^{-3}$ of the mean thickness, the model produces fault spacings that reflect the temporal structure of the forcing. When driven by sea-level-derived signals, the resulting spacing distributions reproduce the observed concentration of power near Milankovitch periods.

These findings demonstrate that plate-thickness variability provides a viable pathway for transmitting sea-level signals into tectonic structure at mid-ocean ridges. At the same time, the model relies on simplified representations of both lithospheric structure and fault mechanics, and does not explicitly resolve the processes by which magma supply variations translate into thickness perturbations. Further work is required to better constrain these processes and to test the mechanism against expanded observational datasets.

More broadly, this study highlights the sensitivity of flexural stress to small variations in plate properties and introduces a framework for coupling external forcing to tectonic pattern formation. This framework may be applicable beyond mid-ocean ridges, including other settings where bending and localisation interact.

\paragraph{Acknowledgements}
The authors thank P.~Howell for advice in formulating the yield-moment-evolution equation, and R.~Buck, S.~Carbotte, J.~Escartin, J.A.~Olive and  J.S.~Wettlaufer for helpful discussions, questions and suggestions.  This research received funding from the European Research Council under Horizon 2020 research and innovation program grant agreement 772255.

\appendix

\section{Reversion to elastic form}
\label{sec:elastic-reversion}

Here we show how the elastic--viscoplastic formulation reverts to the elastic formulation used by \cite{Buck2001}.  Assume that $\ystress\to\infty$ such that $F(M/\yield)=0$ and there is no plastic deformation. Then equation~\eqref{eq:elastic--viscoplastic-constitutive} becomes
\begin{displaymath}
    \ld{M}{t} = -EI\ld{\curvature}{t}.
\end{displaymath}
Assuming a steady state in which the Eulerian time-derivatives are zero and integrating with respect to $x$, we obtain $M(x) = -EI\curvature(x) + c$, where $c$ is a constant of integration. Using the boundary conditions \eqref{eq:bc-isostatic-curvature} and \eqref{eq:bc-zero-moment}, we obtain
\begin{equation}
    \label{eq:app-elastic-moment}
    M(x) = -EI(w'' - \curvature_n),
\end{equation}
where we have expressed the curvature as $\curvature = w''$.  Taking two $x$-derivatives of the vertical force balance \eqref{eq:vertical-force-balance} and eliminating $w''$ using \eqref{eq:app-elastic-moment} gives
\begin{equation}
    M'''' - \Delta\rho g \left(\curvature_n - \frac{M}{EI} + \frac{\pert{h}''}{2}\right) = 0.
\end{equation}
This is a purely elastic, instantaneous equation for $M$ in which $\curvature_n,$ $I$ and $\pert{h}$ are specified functions of $x$. We analyse this equation in Appendix~\ref{sec:linearised-analysis}.

Alternatively, we can use \eqref{eq:app-elastic-moment} to eliminate $M$ from \eqref{eq:vertical-force-balance}. If we then assume that the plate has uniform thickness ($\pert{h}=0$),
\begin{displaymath}
    B w'''' + \Delta\rho g w = 0,
\end{displaymath}
which is in the elastic form used by \cite{Turcotte2002} and \cite{Buck2001}. The bending stiffness $B\equiv EI$ is uniform.

\section{Rescaling to obtain the non-dimensional governing system}
\label{sec:rescaling}

We now gather the equations \eqref{eq:curvature-definition}, \eqref{eq:elastic--viscoplastic-constitutive}, \eqref{eq:yield-weakening}, \eqref{eq:vertical-force-balance} governing variables $w,\,\curvature,\,M,\,\yield$, and the boundary conditions \eqref{eq:bc-far-field}, \eqref{eq:bc-isostatic-height}, \eqref{eq:bc-isostatic-curvature}, \eqref{eq:bc-zero-moment} and rescale symbols to obtain a dimensionless system.  We rescale with
\begin{equation}
    \label{eq:characteristic-scales}
    \begin{gathered}
    [x] = L\equiv \sqrt{2}\ell_0,\quad \quad [t] = L/U,\quad  [h,\Tilde{h}]=h_0,\quad [w]=h_0R,\\ \quad[M,\yield]=EI_0h_0R/L^{2},\quad  [\curvature]=h_0R/L^{2},\quad [I]=I_0 \equiv h_0^3/12,   
    \end{gathered} 
\end{equation}
where $R$ is the ridge-axis density ratio of eq.~\eqref{eq:bc-isostatic-height}, and the characteristic length-scale $L$ is about 1.4$\times$ the flexural--isostatic bending length $\ell$ of equation~\eqref{eq:flexural-isostatic-length}, evaluated for $h=h_0$. After rescaling, the equations become
\begin{subequations}
    \label{eq:dimensionless-governing-equations}
    \begin{gather}
     \label{eq:dimless-gov-force-bal}
     M'' = 4w + 2{h_1}/{R}\quad\text{with }w''=\curvature, \\
     \left[\ld{\,}{t} + \deb\, F(M/\yield)\right]M = -h^3\ld{\curvature}{t},\\
     \ld{\,}{t}\ln{\yield} = -\frac{\deb\,|G(M/\yield)|}{\curvatureref h^3}\left[\yield - (1-f_W)h^2\yieldref\right].
    \end{gather}
\end{subequations}
where $x,\,t,\,h,\,w,\,\curvature,\,M$ and $\yield$ are now dimensionless, $f_W$ is a fraction in the range $0<f<1$, and $h_1\equiv\pert{h}/h_0$ is the dimensionless plate-thickness perturbation. The factor of 2 in eq.~\eqref{eq:dimless-gov-force-bal} arises from the partitioning of thickness perturbations in the load term (eq.~\ref{eq:vertical-force-balance}). In equations~\eqref{eq:dimensionless-governing-equations}, we also introduced the dimensionless quantities
\begin{subequations}
    \begin{align}
        \deb &\equiv \frac{E L}{\eta U},\\
        \curvatureref &\equiv \frac{\weakcurvature}{h_0R/L^2},\\
        \yieldref &\equiv \frac{\ystress h_0^2/6}{EI_0h_0R/L^{2}},\\
        G(m) &\equiv m F(m).
    \end{align}
\end{subequations}
The first of these, the Deborah number, is the ratio of the timescale for elastic stress accumulation by unbending $L/U$ with the Maxwell timescale $\eta/E$ for stress relaxation by viscous flexure above the plastic yield moment. The second, $\curvatureref$, is a dimensionless measure of plastic curvature over which the yield moment decreases from its dimensionless initial value $\yieldref h^2(x,t)$ to its minimum value. This minimum value is given as a fraction $1-f_W$ of that initial value.  Finally, we have defined $G()$ as a modified yield function.

The dimensionless boundary conditions are 
\begin{subequations}
    \label{eq:dimensionless-boundary-conditions}
    \begin{align}
        \label{eq:dimless-bc-w}
        w(0,t) &= h(0,t), \\
        \curvature(0,t) &= \curvature_n(0,t) = \frac{2h(0,t)}{\sqrt{I(0,t)}},\\
        M(0,t) &= 0, \\
        \yield(0,t) &= \yield^\text{max}(0,t) =  \yieldref h^2(0,t),\\
        \label{eq:dimless-bc-farfield}
        \ld{\,\cdot}{t} &= 0 \qquad \text{for }x\to\infty,
    \end{align}
\end{subequations}
where the last condition applies in the far field and applies to any relevant quantity. Rescaling equation~\eqref{eq:plate-thickness-decomp} for the (now dimensionless) plate thickness $h(x,t)$,
\begin{equation}
    \label{eq:append-plate-thick-decomp}
    h(x,t) = 1 + h_1(x,t),
\end{equation}
where, again, $h_1\equiv\pert{h}/h_0$ is the dimensionless perturbation. After rescaling and substitution of \eqref{eq:append-plate-thick-decomp}, the dimensionless second moment of area \eqref{eq:second-moment-area} becomes $I(x,t) = h^3(x,t) = (1+h_1)^3$.

This system of dimensionless governing equations \eqref{eq:dimensionless-governing-equations}, dimensionless boundary conditions \eqref{eq:dimensionless-boundary-conditions} (with dimensionless parameters and yield functions), and dimensionless plate-thickness \eqref{eq:append-plate-thick-decomp} comprise the mathematical problem for analysis. In particular, we seek an understanding of how specified perturbations $h_1(x,t)$ affect the occurrence and spacing of viscoplastic kinks in the plate.

\section{Linearised analysis of elastic flexure}
\label{sec:linearised-analysis}

Here we seek the variations in bending moment $M$ that arise from the perturbation in plate thickness. We formally assume harmonic plate-thickness perturbations that satisfy $|h_1| \sim \epsilon \ll 1.$ This enables a linearisation of the problem around a base state that is exactly the solution of \cite{Buck2001}. The analysis begins in the usual way with expansion of variables into Taylor series and truncation at first order. For example, $M(x,t) \sim M_0(x) + M_1(x,t)$, where $|M_0|$ is assumed to be $O(1)$ and $|M_1|$ is $O(\epsilon)$. We substitute these expansions and neglect terms of $O(\epsilon^2)$ or smaller.  Expanding and truncating the second moment of area gives $I(x,t) \sim 1 + 3h_1$ and the neutral curvature becomes $\curvature_n(x,t) \sim 2 - h_1$. 

\begin{figure}
    \centering
    \includegraphics[width=\linewidth]{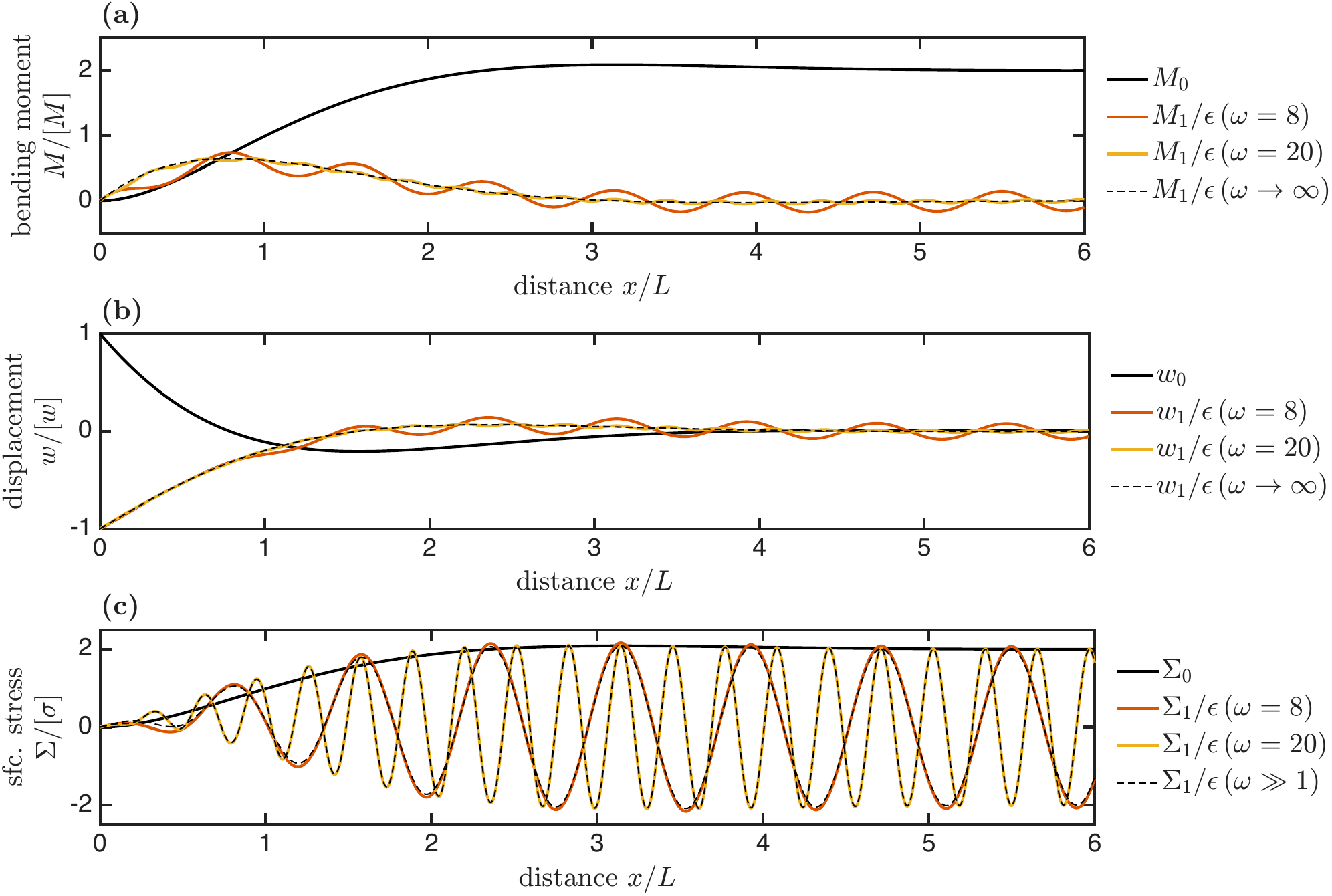}
    \caption{Results from the linearised analysis of elastic unbending plotted at $t=\pi/\omega$. The first-order (perturbation) curves are plotted for $\omega=8$ and $\omega=20$. Asymptotic results are plotted as dashed curves. \textbf{(a)} Bending moment $M_0$ and $M_1$. \textbf{(b)} Vertical displacement $w_0$ and $w_1$ computed from $M$ using eqn.~\eqref{eq:dimless-gov-force-bal}. The asymptotic curve is given by $w_1\sim h_1(0,t)\e^{-x}\cos x$. \textbf{(c)} Fibre stress $\Sigma_0$ and $\Sigma_1$ evaluated at the plate's upper surface.}
    \label{fig:linearised-analysis}
\end{figure}

We analyse the non-dimensional version of the elastic moment equation \eqref{eq:app-elastic-moment} after rescaling with \eqref{eq:characteristic-scales},
\begin{equation}
    \frac{M''''}{4} + \frac{M}{h^3} = \curvature_n + \frac{h_1''}{2R}.
\end{equation}
Substituting our truncated Taylor series and balancing terms at each order we obtain the coupled equations
\begin{subequations}
    \begin{align}
        \label{eq:M0-equation}
        M_0''''/4 + M_0 &= 2,\\
        \label{eq:M1-equation}
        M_1''''/4 + M_1 &= 3h_1M_0 - h_1 + h_1''/2R.
    \end{align}
\end{subequations}
At leading order the moment is induced by unbending from the neutral curvature (2 in dimensionless units) to zero curvature. This leads to a change in $M_0$ from zero at the axis to 2 in the far field, along the curve \citep{Buck2001}
\begin{equation}
    \label{eq:moment-base-state}
    M_0(x) = 2\left[1 - \e^{-x}\left(\cos x + \sin x\right)\right].
\end{equation}
We have used boundary conditions \eqref{eq:dimless-bc-w} and \eqref{eq:dimless-bc-farfield}. The former, substituted into \eqref{eq:dimless-gov-force-bal}, implies $M_0''(0,t)=4$. The solution \eqref{eq:moment-base-state}, plotted in Fig.~\ref{fig:linearised-analysis}a, goes through a maximum of $M_0(\pi) = 2(1+\e^{-\pi})\approx2.1$.

An analytical solution to \eqref{eq:M1-equation} is available for the particular choice
\begin{equation}
    \label{eq:dimless-sinusoidal-perturbation}
    h_1(x,t) = \epsilon\,\e^{i\omega(x-t)},
\end{equation}
where $\epsilon \ll 1$ is a constant, dimensionless amplitude and $\omega = 2\pi L/(UT)$ is a dimensionless, angular frequency. This frequency corresponds to a mode of sea-level forcing with dimensional period $T$ (e.g., the $T=41$-ky oscillation).  Because the dimensional wavelength of the perturbation, generated by plate spreading at half-rate $U$, is $\lambda = UT$, the dimensionless wavenumber is also $\omega$. 

The solution $M_1(x,t)$ must be finite for $x\to\infty$, and must satisfy conditions $M_1(0,t)=0$ and $M_1''(0,t) = (4+2/R)h_1(0,t)$. The latter condition arises by combining \eqref{eq:dimless-bc-w} with \eqref{eq:dimless-gov-force-bal} at $O(\epsilon)$.  Substituting the perturbation \eqref{eq:dimless-sinusoidal-perturbation} and the base state \eqref{eq:moment-base-state} into \eqref{eq:M1-equation} and using \textsc{Matlab}'s symbolic toolbox \citep{code-repo},
\begin{equation}
    \label{eq:moment-perturbation}
    M_1(x,t) = 
    \e^{-x}\left\{h_1(0,t)\left[A_\omega\cos x - B_\omega\sin x\right] 
    + h_1(x,t)\left[C_\omega\e^{-ix} - D_\omega\e^{ix}\right]\right\} + h_1(x,t)\mathcal{F}_\omega,
\end{equation}
where the frequency-dependent coefficients are
\begin{subequations}
    \small
    \begin{align}
        A_\omega &= \frac{\omega^{10}/R + 2\omega^8 + (48/R + 240)\omega^6 - 960i\omega^5 - 1548\omega^4 + (960 - 1024/R)\omega^2 - 4840i\omega + 5632}{\omega^{12}/2 + 26\omega^8 - 416\omega^4 - 2048},\\
        B_\omega &= \frac{\omega^{11} - 5\omega^9 + (2/R + 36)\omega^7 + 96i\omega^6 - (8/R + 72)\omega^5 - 384i\omega^4 + (128/R - 1072)\omega^3 + 384i\omega^2 - (512/R + 1408)\omega - 1536i}{\omega^{11}/2 - 2\omega^9 + 34\omega^7 - 136\omega^5 + 128\omega^3 - 512\omega},\\
        C_\omega &= \frac{12\left[(1 - i)\omega-4\right]}{i\omega^5 - (2 + 6i)\omega^4 + (12 + 16i)\omega^3 - (32 + 16i)\omega^2 + 32\omega},\\
        D_\omega &= \frac{12\left[(1 + i)\omega + 4\right]}{i\omega^5 - (2 - 6i)\omega^4 - (12 - 16i)\omega^3 - (32 - 16i)\omega^2 - 32\omega},\\
        \label{eq:Fom_unapproximated}
        \mathcal{F}_\omega &= \frac{24}{\omega^4 + 4} - \frac{2\omega^2}{R(\omega^4 + 4)} - \frac{4}{\omega^4 + 4}.
    \end{align}
    \normalsize
\end{subequations}
Equation~\eqref{eq:moment-perturbation} is plotted in Fig.~\ref{fig:linearised-analysis}a for two values of $\omega$.

The first thing to notice about the solution for $M_1$ in equation~\eqref{eq:moment-perturbation} is that it has a part that decays with $\e^{-x}$ and another part that doesn't decay with $x$.  We refer to the decaying part as the near-field contribution and the non-decaying part as the far-field contribution.

The near-field contribution has two parts.  The first, associated with the homogeneous solution to \eqref{eq:moment-perturbation}, represents moment perturbations that arise from vertical displacements of the ridge axis, $h_1(0,t)$.  As this height oscillates, it causes flexure that varies on the flexural--isostatic length scale. The second contribution is part of the particular solution forced by $3h_1M_0$. It is due to variations in the second moment of area $I_1(x,t)$ interacting with base-state flexure to generate moment variations on the wavelength of the perturbation.

The far-field contribution has three parts, which are separated in the expression of $\mathcal{F}_\omega$ in eq.~\eqref{eq:Fom_unapproximated}. The first is the non-decaying contribution of $I_1(x,t)$ interacting with base-state flexure; the second is due to perturbations of the ingrown curvature; the third is due to the variable load associated with thickness perturbations.

In assessing the amplitude of these contributions, we consider how the coefficients in \eqref{eq:moment-perturbation} scale at large $\omega$.  At leading order,
\begin{equation}
    \label{eq:large-omega-coefs}
    A_\omega \sim \frac{2}{\omega^2 R},\quad
    B_\omega \sim 2,\quad
    C_\omega \sim \frac{12(i+1)}{\omega^4},\quad
    D_\omega \sim \frac{12(i-1)}{\omega^4},\quad
    \mathcal{F}_\omega \sim - \frac{2}{\omega^2R}, \qquad\text{for $\omega\gg1$}.
\end{equation}
Hence we see that at large $\omega$, the leading contribution to the bending moment $M$ \textit{at the perturbation wavelength} comes from the neutral curvature variation in $\mathcal{F}_\omega$, and is $O(\epsilon/\omega^2)\ll\epsilon$. Perturbations to $M$ at the wavenumber $\omega$ are thus negligible.

These asymptotics also indicate that for large $\omega$ (that are relevant for our purposes), the term in $B_\omega$ makes the dominant contribution; asymptotically,
\begin{equation}
    \label{eq:moment-perturbation-highfreq}
    M_1(x,t)\sim -2h_1(0,t)\e^{-x}\sin x, \qquad\text{for $\omega\to\infty$}.
\end{equation}
This result shows that the largest of the $O(\epsilon)$ contributions to the moment varies on the flexural--isostatic wavelength (not the perturbation wavelength). Therefore it cannot contribute to pacing faults.

Normal faults are caused by extensional fibre stresses $\sigma \equiv \sigma_{xx}$ in the plate.  The flexural contribution to these stresses varies linearly in $z$ away from the mid-plane of the plate,
\begin{equation}
    \sigma(x,z,t) = z{M}/{h^3},
\end{equation}
where we have rescaled $z$ with $h_0$ and stress with $Eh_0^2R/L^2$. Faults initiate at the upper surface of the plate, and hence evaluating at $z = h(x,t)/2$ we have 
\begin{equation}
    \Sigma \equiv \sigma(x,h/2,t) = M/2h^2.
\end{equation}
The leading-order surface stress due to the unperturbed flexure is $\Sigma_0(x) = M_0(x)/2$; this stress was invoked by \cite{Buck2001} in explaining the distribution of normal-fault throws with $x$ at the East Pacific Rise. More importantly in the present context, at first order we have
\begin{equation}
    \label{eq:surface-stress-linearised}
    \Sigma_1(x,t) = M_1/2 - h_1 M_0.
\end{equation}
This equation is plotted in Fig.~\ref{fig:linearised-analysis}a for two values of $\omega$. The first term on the right-hand side inherits its large--$\omega$ behaviour from \eqref{eq:large-omega-coefs}, and hence carries a factor of $\omega^{-2}$ compared with the second term. The second term, arising from base-state unbending acting on variations of bending stiffness (due to plate thickness), has the following important properties: it is of order $\epsilon$, has the wavelength of the plate-thickness perturbation $h_1$, is not diminished at large $\omega$, and has an amplitude that increases to saturation on the flexural--isostatic length-scale over which normal faulting develops. \textit{We therefore anticipate that this stress perturbation has the potential to pace faulting.}

Expanding the surface stress \eqref{eq:surface-stress-linearised} in the limit of large $\omega$ using \eqref{eq:moment-base-state} and \eqref{eq:moment-perturbation-highfreq}, we find
\begin{equation}
    \Sigma_1(x,t) \sim -h_1(0,t)\e^{-x}\sin x - 2h_1(x,t)\left[1 - \e^{-x}\left(\cos x + \sin x\right)\right] \qquad\text{for $\omega\gg1$}.
\end{equation}
To gain a crucial insight, we can interpret this result in light of the tension-positive sign convention for stress. Recalling that $M_0>0$, the second term states that where the plate is thinner (i.e., where $h_1$ is negative), the flexural fibre stress at the plate's upper surface is more tensile. \textit{We therefore anticipate faults may preferentially appear where the plate is thinner.}

\section{Numerical methods}
\label{sec:numerical}

The full model with elastic--viscoplastic rheological law must be solved numerically.  We do this in terms of four fields $w,\,\curvature,\,M,\,\yield$ that evolve in space and time, constrained by dimensionless equations \eqref{eq:dimensionless-governing-equations} with conditions \eqref{eq:dimensionless-boundary-conditions}.  Variables are discretised on a uniform grid with dimensionless spacings $\delta_x$ and $\delta_t$. Spatial derivatives in equations~\eqref{eq:dimless-gov-force-bal} are discretised by a standard 3-point stencil with second-order accuracy.

Lagrangian derivatives are discretised with a semi-Lagrangian method \citep[e.g.,][]{spiegelman2006semi}. The characteristics are straight lines because the dimensionless advective velocity is $\xhat$. Interpolation at the foot of the characteristic at the previous time-step is by monotone cubic interpolation \citep{Fritsch1980}.

The model evolution is initiated with a spin-up phase, where the solution is smoothly forced away from the trivial solution.  This is achieved by a continuation function, 
\begin{equation}
    C(x,t) = \frac{1}{2}\left[1-\tanh\left(\frac{X-X_0}{\delta_X}\right)\right],
\end{equation}
where $X=x-t$ and $X_0,\delta_X$ are constants. We use $X_0=-1$ and $\delta_X = 1/3$ such that $C(0,0)\approx 0$ and $C$ ramps up with time from there. We apply this function to the axial height $w(0,t)$, the neutral curvature $\curvature(0,t)$ and to the perturbation amplitude $\epsilon$.

The code is written in the framework of the Portable Extensible Toolkit for Scientific Computation \citep[PETSc;][]{petsc-efficient, petsc-user-ref}.  It uses a standard Newton method for non-linear systems with an LU-preconditioned Krylov (GMRES) linear solver. All four governing PDEs are bundled into the same Newton solver.  In rare cases, this solver does not reduce the norm of the nonlinear residual to below an absolute tolerance of $10^{-7}$ after 10 Newton steps. In such cases, $\delta_t$ is reduced and the step is re-attempted.  The CFL number, $\delta_t/\delta_x$, is required to be a unit fraction: $1/n$ for integer values of $n$. After a failed step, $n$ is incremented by one; after 100 successful steps, $n$ is decremented by one.  Figures published here are generated with $\delta_x = 10^{-3}$ or smaller.

The code version used for the calculations in this paper is available in an online repository \citep{code-repo}.

\section{Parametric sensitivity}
\label{sec:sensitivity}

The model results depend on parameter choices that, in some cases, are not well constrained.  Moreover, if parameters fall outside a certain range, the model cannot capture the observations even qualitatively.  For example, for yield moments greater than the maximum moment achieved by elastic unbending of the plate, no plastic failure occurs (i.e., no plastic kinks), indicating that our theory cannot explain observations.  While possible, this is not an outcome that requires detailed study.  

\begin{figure}
    \centering
    \includegraphics[width=\linewidth]{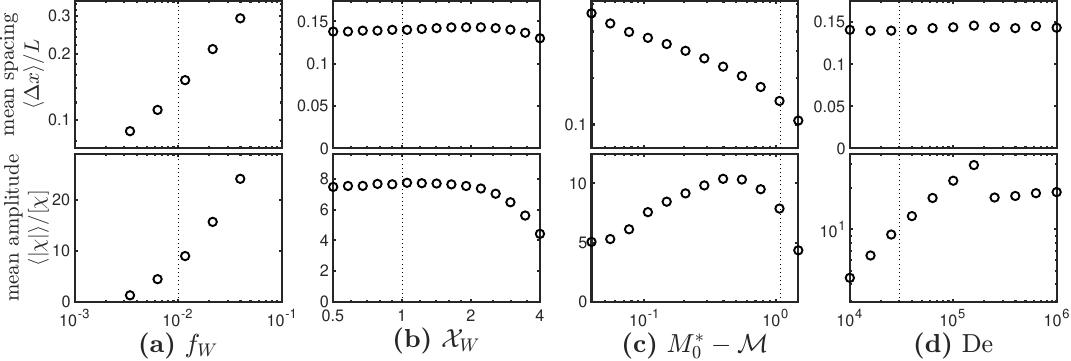}
    \caption{Mean spacing (top row) and amplitude (bottom row) of kink-faults in a uniform-thickness plate, as a function of individual dimensionless parameters. All parameters are held at their reference values (Tab.~\ref{tab:model-parameters}) except the one on the $x$ axis; dotted lines mark the reference value of that parameter. \textbf{(a)} Weakening magnitude $f_W$. \textbf{(b)} Plastic-strain scale for weakening $\curvatureref$. \textbf{(c)} Reference yield moment compared with the maximum dimensionless elastic moment $M_0^* = 2(1+\e^{-\pi}).$  \textbf{(d)} The Deborah number.}
    \label{fig:sensitivity}
\end{figure}

Here we briefly consider variations of four dimensionless parameters around their preferred values, given in Table~\ref{tab:model-parameters}.  The results are summarised in Figure~\ref{fig:sensitivity} in terms of the predicted kink-fault spacing and amplitude of kink curvature. The latter is obtained by finding the maximum absolute curvature for each kink in the plate and then taking the average of these values.  In these plots, each point represents a model run with all parameters at their reference values except the one shown on the $x$ axis, which varies between runs.  The reference value of the parameter on the $x$ axis is indicated by a vertical dotted line.

Panels (a) of Fig.~\ref{fig:sensitivity} show the response to varying the maximum fraction of plastic weakening, $f_W$. This is the maximum relative drop in the yield moment at a point, with progressive plastic strain.  A larger value of $f_W$ gives kinks greater self-weakening.  With increasing $f_W$, the curvature of kinks increases as their spacing increases.  The overall flattening of the plate is achieved through fewer but higher amplitude kinks.  We could think of this as a trade-off between normal-fault spacing and throw.

Panels~(b) show the response to the plastic-strain scale $\weakcurvature$ for weakening.  This is the range of plastic strains over which the yield moment drops by $f_W$.  Evidently, these results are insensitive to this parameter except where it reaches 2--4$\times$ its reference value. This is because absolute curvature of kinks readily exceeds $\weakcurvature$, leading to saturated weakening in kinks.

Panels (c) show the response to changes in the reference yield moment $\yieldref$.  Recall that for a plate of uniform thickness, the initial yield moment $\yield(0,t) = \yieldref,$ prior to any plastic deformation.  To clarify the meaning of variations in this parameter, we plot it in terms of $M_0^* - \yieldref$, where $M_0$ is the maximum value of the elastic bending moment in the absence of plastic failure (eq.~\eqref{eq:moment-base-state}). For a uniform plate with $\yieldref>M_0^*$, there are no plastic kinks whatsoever.  Hence for $(M_0^* - \yieldref) \to 0$ from above, we expect $\Delta x\to\infty$.  This is what is observed in upper-panel~$(c)$. The effect of $\yieldref$ on the amplitude of kinks is non-monotonic, but the sensitivity at $\yieldref=1$ is positive: increasing $\yieldref$ gives larger and more widely spaced kinks.

Panel (d) shows the effect of the Deborah number. Increasing $\deb$ corresponds to decreasing the viscosity of plastic deformation.  The mean spacing of kinks is insensitive to $\deb$, but the amplitude increases with $\deb$. This is because kinks become more narrowly localised, such that the increase in amplitude compensates the decrease in width to conserve the integral of $\curvature$ across the kink. A change in the trend in lower-panel~(d) is evident at about $2\times10^5$. Above this value of $\deb$, the progression of kinks becomes disordered and aperiodic.

\begin{figure}
    \centering
    \includegraphics[width=0.8\linewidth]{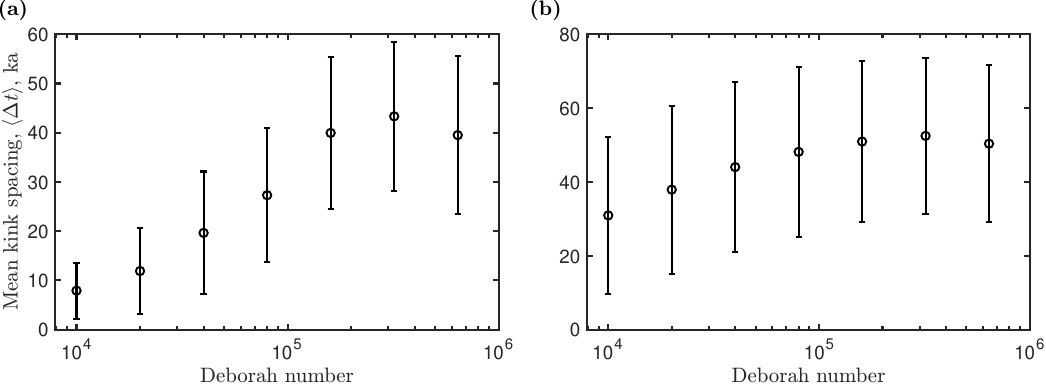}
    \caption{Mean kink spacing as a function of Deborah number for non-uniform plate thickness with $\epsilon=0.01$. Error bars show one standard deviation. \textbf{(a)} For plate thickness perturbation in proportion to a random time-series $N(t)$, as described in section \ref{sec:results-sea-level}. \textbf{(b)} For plate thickness perturbation in proportion to sea-level rate $\dot{S}(t)$.}
    \label{fig:deborah-sensitivity-noise}
\end{figure}

The sensitivity to Deborah number differs when $h(x,t)$ is non-uniform.  Figure~\ref{fig:deborah-sensitivity-noise} shows the mean kink spacing cast in terms of dimensional time as a function of $\deb$. Panel~(a) uses the random time-series $N(t)$, which is uniformly distributed in $[-\sqrt{3},\,\sqrt{3}]$.  The trend is evident and with $\deb\approx 2\times10^5$, a mean spacing of about 40~ka is attained.  Panel~(b) uses the sea-level rate $\dot{S}(t)$.  A similar trend appears, but it is muted relative to the case in panel~(a).

\section{Pleistocene sea-level record and its extension}
\label{sec:sea-level-extension}

The reconstruction of $\delta^{18}$O by \cite{Huybers2007} spans 2.58~Ma to the present with a sampling interval of 1~kyr. To extend it, we first identify two segments to serve as templates.  The template for the early Pleistocene is the period from 2.58 to 2~Ma; this represents the ``41-kyr world.'' The template for the late Pleistocene is from 1 Ma to present and represents the ``100-kyr world.''  

Each of these templates is then fed to the algorithm proposed by \cite{schreiber1996improved} to generate synthetic records. This creates the synthetic time-series and iterates its array of Fourier phases. The algorithm is complete when the synthetic series matches \textit{(i)} the power spectrum of the template series, and \textit{(ii)} the amplitude distribution of the template series in terms of its rank-ordered entries.  

Length mismatches (the output segments are longer than their templates) are handled by interpolating the target amplitude spectrum onto the output-length frequency grid, and by stretching the sorted template values to the output length. Each synthetic segment is therefore a random realisation that shares both the spectral character and the marginal distribution (range, skewness) of its template, but is otherwise independent of the template series.

The two synthetic segments are concatenated with the original series to form the extended series, which spans 10~Ma to 7.5 million years in the future. 

The code used for this task, which was generated by the AI system Claude, is provided in the online repository \citep{code-repo}.

\setstretch{1.0}
\bibliographystyle{abbrvnat}
\bibliography{references}
\end{document}